\documentclass[%
preprint,
bibnotes,
 amsmath,amssymb,
 aps,
]{revtex4-1}

\usepackage[sort&compress]{natbib}
\usepackage{graphicx}
\usepackage{dcolumn}
\usepackage{bm}

\begin{document}

\preprint{preprint}

\title{Modulation instability in noninstantaneous Kerr media with walk-off effects and cross-phase modulation for two copropagating beams experiencing anomalous GVD regime}

\author{Askery Canabarro}
 \affiliation{International Institute of Physics, Federal University of Rio Grande do Norte, 59070-405 Natal, Brazil}
\affiliation{Grupo de F\'isica da Mat\'eria Condensada, N\'ucleo de Ci\^encias Exatas, Campus Arapiraca, Universidade Federal de Alagoas, 57309-005, Arapiraca-AL, Brazil}
 \email{askery.canabarro@arapiraca.ufal.br}

\author{B. Santos}
 \affiliation{Observat\'orio Nacional, 20921-400, Rio de Janeiro - RJ, Brazil}

\author{B. de Lima Bernardo}
 \affiliation{Departamento de F\'isica, Universidade Federal da Para\'iba, Caixa Postal 5008, 58059-900 Jo\~ao Pessoa, PB, Brazil}

\date{\today}

\begin{abstract}
The dynamics of the modulation instability induced by cross phase modulation is studied by considering the influence of the walk-off and noninstantaneous response effects for two copropagating optical fields travelling in the anomalous regime of dispersion. To do so, we make use of extensions of the nonlinear Schrödinger equation jointly with the Debye model for polarization, which is shown to be effective and simplified when compared to the implementation of other methods for the noninstantaneous nonlinear response. In analyzing the sideband formation, two bands are observed with different behaviors with respect to the way the maximum gain and the respective frequency vary with increasing phase mismatch. Further, we also show that the manner in which the maximum gain as well as its corresponding frequency scale with the delay parameter $\tau$ is substantially different from the cases of both fields experiencing the normal group-velocity dispersion regime and the case of mixed regimes. These facts may give rise to many new possibilities for the evolution of the modulated wave when compared to the nonanomalous cases studied in the literature. 

\begin{description}
\item[Keywords]
Fiber Optics; Instability and chaos.
\end{description}
\end{abstract}

\maketitle


\section{Introduction}

Modulation instability (MI) was a phenomenon firstly detected in hydrodynamic systems whereby water wave trains were found to disintegrate into irregular structures \cite{benjamin}, being nowadays possibly associated with the underlying physical mechanisms responsible for the generation of soliton as well as rogue waves \cite{Agrawal:11,Janssen,dysthe,onoratoPRL,onoratoPRL2,Solli2007,priya,stepiPRA}. Moreover, its manifestation is now known to be commonly found in a myriad of nonlinear physical systems as a result of intensity-dependent perturbations due to the strong nonlinear effects that take place in nature. In this form, besides fluid systems, MI has also been detected in optical systems \cite{Solli2007}, Bose-Einstein condensates \cite{everitt},  strongly magnetized plasma \cite{litvak,wuPRL}, electromagnetic transmission lines \cite{ZAKHAROV2009540}, just to mention some examples. However, the effect, formation and applications of MI have been more prominently investigated in the context of light propagation, being the subject of intense studies in the optical community for decades. In this regard, the phenomenon is characterized by the transformation of long optical signals into pulse trains \cite{agrawal1,Agrawal:11} rendering sidebands in the spectrum accompanied by modulation growth and the generation of a group of isolated waves separated in space or time \cite{kiv}.

The time evolution of many systems subjected to MI is generally described by the one-dimensional nonlinear Schrödinger equation (NLSE), which gives the possibility of modeling the dynamics of weakly nonlinear dispersive waves propagation in most of the system cited above \cite{biodini}. In the specific case of optical systems, it is well-known that MI manifests itself in a cw light waves when propagating in media with anomalous dispersion; a consequence of the so-called self-phase modulation (SPM)\cite{moll}. However, it is also known that in the case of two or more copropagating optical fields inside a nonlinear material, which is referred to as cross-phase modulation (XPM), MI can also appear both in the anomalous and normal dispersion regimes, which provides a much richer scenario of study \cite{agrawal2}. Indeed, MI induced by XPM has been extensively studied in the literature \cite{chow1,Zhang14,nithy1,BISWAS20143299,nithy2,Xiang11,ZHOU20091440,Zhong14,ZHONG2007271,daSilva:09,Tanemura03,PhysRevE.74.036614,Vasantha10,edmunson04,lin04,rothen}, including by the authors \cite{canabarro0,canabarro1,canabarro2,canabarro3}. One of main motivations of this study is the fact that XPM is a practical mechanism of wavelength conversion and optical switching \cite{harris}. Nevertheless, more sophisticated technologies have appeared recently based on the XPM properties. For example, in the implementation of quantum nondemolition measurements and quantum logic gates \cite{Schmidt96,Fushman07}, with improvements made in the context of weak measurements \cite{Hallaji15}.    

With the advancement and precision requirements of the XPM applications, it is increasingly necessary to model the dynamics of XPM taking into account the influence of many perturbations involved, apart from the predominant nonlinear and dispersive effects \cite{Agrawal:11,ZAKHAROV2009540}. Among the many types of perturbation, some of the most reported in the literature are the walk-off effect, in which the beam propagation direction tends to deviate from the direction of the wave vector in anisotropic media \cite{canabarro0}, and the non-instantaneous optical response of the Kerr nonlinearity \cite{canabarro0,canabarro1,canabarro2,canabarro3,NITHY3,chou09,contiPRL,Vasantha10,akhmediev98}. Therefore, given the vast possibility of propagating media, perturbation effects, and dispersion regimes for the copropagating beams influencing the XPM dynamics, it is essential to realize numerical tests with of such possibilities in order to better implement the specific type of propagating media and optical fields to be used in applications. 

In this article, we study MI in a XPM system for the case of two optical fields copropagating in a nonlinear kerr medium in which both the walk-off and noninstantaneous effects are taken into account. In doing so, it is considered that both fields travel through the medium in the anomalous regime of dispersion. As a matter of fact, the noninstantanous effect, which is generally associated with the delayed nonlinear rearrangement of charges of the material in the presence of intense light waves, is possibly the least understood element affecting the dynamics of MI in optical systems. In fact, in the literature there is only a limited amount of works reporting the influence of non-instantaneity in the present context. Here, we account for such an effect in a rather simplified form by using the Debye model to describe the polarization \cite{carbou2006,Cambournac02}, in place of the non-trivial treatement with integro-differential equations usually applied. 

Our results for the sideband generation show that, contrary to the normal and mixed regimes \cite{canabarro0,canabarro2}, two bands take place for a large range of the phase mismatch between the optical fields, each with very specific behaviors. We also demonstrate that the maximum gain and the respective frequency scale with the finite response time in a different fashion with respect to the scenarios in which both fields experiencing the normal group-velocity dispersion regime and also the case of mixed regimes. These new facts uniquely observed in the case analyzed in this present work may be used for the creation of potential more favorable habits for the appearance of solitary as well as rogue waves due to the richer formation of MI sidebands. 

\section{Theoretical model and linear stability analysis}
\label{sec:model}

To start with we analyze the case in which two optical pulses of distinct frequencies and same polarizations propagates on a single-mode optical fiber
with Kerr nonlinearity. This can be described by the set of coupled nonlinear 
Schrödinger equations (CNLSEs):

\begin{eqnarray}
\label{eq1}
& &\imath \frac { \partial E_{1} } {\partial z} + \frac{\imath}{\nu_{g1}} \frac{\partial E_{1} } {\partial t} =
 \frac{\beta_1}{2} \frac{\partial^2 E_{1} } {\partial t^2}- \gamma_1 (|E_1|^2 + 2|E_2|^2) E_1, \nonumber \\
& &\imath \frac { \partial E_{2} } {\partial z} + \frac{\imath}{\nu_{g2}} \frac{\partial E_{2} } {\partial t} =
 \frac{\beta_2}{2} \frac{\partial^2 E_{2} } {\partial t^2}- \gamma_2 (|E_2|^2 + 2|E_1|^2) E_2, \nonumber \\
\end{eqnarray}
where $E_i = E_i(t,z)$ corresponds to the electric field envelope of the wave which  propagates along the $z$ axis with a group velocity $\nu_{gi}$ 
in a retarded time frame $t = (T - z/\nu_g)$. The parameter $\nu_{gi}$ $(i=1,2)$ accounts for the group velocity, $\beta_i$ represents the group
velocity dispersion (GVD) coefficient, and  $\gamma_i = $ is the Kerr coefficient 
experienced by the beam $i$ when propagating through the fiber. The group-velocity mismatch (GVM) is identified by $\delta$ and is given by $\delta = | \nu^{-1}_{g1} - \nu^{-1}_{g2} |$. Moreover, since the two beams have the same polarization, it is necessary the introduction of the factor 2 before the XPM terms \cite{Agrawal:11,ZAKHAROV2009540,kiv,biodini,moll,agrawal1,agrawal2}.

In writing the equations above we implicitly considered that the nonlinear response of the medium is instantaneous. However, in dealing with the propagation of ultrashort pulses, in addition to the instantaneous response of the electronic contribution, the thermal effects as well as the re-orientational nonlinearity contributes to a slow response of the medium whose time scale ranges from picoseconds to nanoseconds. It is understood that the vibrational Raman effect is the predominant factor for the existence of such delayed effects. Due to its simplicity, the Debye relaxational model is adopted, being recognized to be applicable independent of the frequency regime. In this case, the relaxation coefficient has to do with the finite response time, which is a characteristic of the nonlinear material.

In the case of the ordinary Kerr response, the medium presents an instantaneous response as follows

\begin{eqnarray}
\label{eq2}
P \propto |E|^2 E,
\end{eqnarray}
where $P$ is the nonlinear polarization. In the Debye relaxation model \cite{carbou2006,Cambournac02,debye2,Ziolkowski:93}, the medium has a finite response time $\tau$. This fact is considered if we write $P \propto \chi(E,t) E$. The rate at which $\chi$ changes with time is given by:

\begin{eqnarray}
\label{eq4}
\frac{\partial \chi}{\partial t}= - \frac{1}{\tau} \chi + \frac{1}{\tau} |E|^2.
\end{eqnarray}
This result shows the exponential relaxation of the nonlinear contribution to the stationary solution. In this form, we can take the  time dependent nonlinear response into account in the system of CNLSEs, which are given by \cite{canabarro0,canabarro1,canabarro2,canabarro3,NITHY3,carbou2006,Cambournac02,Potasek87,LIU20082907,Trillo89}

\begin{eqnarray}
\label{eq5}
& &\imath \frac { \partial E_{1} } {\partial z} + \frac{\imath}{\nu_{g1}} \frac{\partial E_{1} } {\partial t} =
 \frac{\beta_1}{2} \frac{\partial^2 E_{1} } {\partial t^2}- \gamma_1 N_1 E_1, \nonumber \\
& &\imath \frac { \partial E_{2} } {\partial z} + \frac{\imath}{\nu_{g2}} \frac{\partial E_{2} } {\partial t} =
 \frac{\beta_2}{2} \frac{\partial^2 E_{2} } {\partial t^2}- \gamma_2 N_2 E_2, \nonumber \\
& &\frac{\partial N_1}{\partial t} = \frac{1}{\tau} (- N_1 + |E_1|^2 + 2|E_2|^2), \nonumber \\
& &\frac{\partial N_2}{\partial t} = \frac{1}{\tau} (- N_2 + 2|E_1|^2 + |E_2|^2),
\end{eqnarray}
where $N = N(z,t)$ is the nonlinear index of the medium. The parameter $\tau$ represents the finite response time. Accordingly, it can be readily seen that it attains the usual Kerr response if we make $\tau \rightarrow 0$. Also, this is applicable even when the slowly varying envelope approximation is not valid \cite{carbou2006}. We shall see that the slow and fast regimes of response times (large and small values of $\tau$, respectively) produces different results.

To analyze the effect of the small harmonic perturbations on the steady-state solution of the dynamical equations shown above, we employ the linear stability analysis. For the situation of two continuous or quasi-continuous optical fields, we assume that the derivatives of Eq. (\ref{eq5}) with respect to time are null. 
%
%
In order to study the stability of the steady-state solution, we consider that noises couple with the main signal as

\begin{eqnarray}
\label{eq6}
 E_{1} &=& [E^0_{1} + e_1(z,t) ] e^{[\imath \gamma_1(|E^0_{1}|^2 + 2|E^0_{2}|^2)z]}, \nonumber \\
 E_{2} &=& [E^0_{2} + e_2(z,t) ] e^{[\imath \gamma_2(2|E^0_{1}|^2 + |E^0_{2}|^2)z]}, \nonumber \\
 N_{1} &=& n_1(z,t) + |E^0_{1}|^2 + 2|E^0_{2}|^2, \nonumber \\
 N_{2} &=& n_2(z,t) + 2|E^0_{1}|^2 + |E^0_{2}|^2,
\end{eqnarray}
where $e_j (z,t)$ is the weak perturbation satisfying $|e_j (z,t)|^2 \ll |E^0_{j}|^2$, and $n_j(z,t)$ as a small perturbation to the stationary nonlinearity.

In addition, from Eqs. (\ref{eq6}) and (\ref{eq5}), we can derive the linearized equations satisfying the perturbations $e_j(z,t)$ and $n_j(z,t)$, which gives us that

\begin{eqnarray}
\label{eq7}
& &\imath \frac { \partial e_{1} } {\partial z} + \frac{\imath}{v_{g1}} \frac{\partial e_{1} } {\partial t} =
 \frac{1}{2} \beta_1 \frac{\partial^2 e_{1} } {\partial t^2}- \gamma_1 n_1 E^0_{1}, \nonumber \\
& &\imath \frac { \partial e_{2} } {\partial z} + \frac{\imath}{v_{g2}} \frac{\partial e_{2} } {\partial t} =
 \frac{1}{2} \beta_2 \frac{\partial^2 e_{2} } {\partial t^2}- \gamma_2 n_2 E^0_2, \nonumber \\
& &\frac{\partial n_1}{\partial t} = \frac{1}{\tau} [- n_1 + E^0_1(e_1+e^{*}_1) + 2E^0_2(e_2+e^{*}_2)], \nonumber \\
& &\frac{\partial n_2}{\partial t} = \frac{1}{\tau} [- n_2 + 2E^0_1(e_1+e^{*}_1) + E^0_2(e_2+e^{*}_2)].
\end{eqnarray}

The solution for this system of coupled complex linear equations can be simply obtained in the Fourier space. Therefore, by writing the Fourier decomposition of the perturbations as
 \begin{eqnarray}
 \label{eq8}
 & & e_j(z,t) = \frac{1}{\sqrt{2\pi}} \int e^{-\imath kz} e^{\imath \Omega t} \widehat{e}_j(\Omega,k) dk d\Omega ,\nonumber \\
 & & n_j(z,t) = \frac{1}{\sqrt{2\pi}} \int e^{-\imath kz} e^{\imath \Omega t} \widehat{n}_j(\Omega,k) dk d\Omega ,\nonumber \\
 \end{eqnarray}
simple expressions for the Fourier components can be found. Here, $k$ is the wave number, and $\Omega$ is the modulation frequency. Next, we investigate how the presence of XPM and non-instantaneous response affect the case in which modulations with equal wave vectors are associated to each beam. As can be seen in Ref. \cite{Tanemura03}, there exist a form to generalize this case. In eliminating $\widehat{n}_j(\Omega,k)$, the amplitude of the perturbation fields harmonic components are found to be coupled as  

\begin{eqnarray}
\label{eq9}
 \left( \right. k &-& \frac{\Omega}{v_{g1}} 
+ \beta_1 \frac{\Omega^2}{2} \left. \right) \widehat{e}_1(\Omega,k)  \nonumber \\
 &+& \frac{\gamma_1 E^0_1}{(\imath \Omega \tau + 1)} \{ E_1^0 [\widehat{e}_1(\Omega,k)+ \widehat{e}^*_1(-\Omega,-k)] \nonumber \\
&-& 2 E_2^0 [\widehat{e}_2(\Omega,k)+ \widehat{e}^*_2(-\Omega,-k)] \}  = 0, \nonumber \\
&~& \nonumber \\
 \left( \right. k &-& \frac{\Omega}{v_{g2}} 
+ \beta_2 \frac{\Omega^2}{2} \left. \right) \widehat{e}_2(\Omega,k) \nonumber \\ &+& 
 \frac{\gamma_2 E^0_2}{(\imath \Omega \tau + 1)} \{ 2E_1^0 [\widehat{e}_1(\Omega,k)+ \widehat{e}^*_1(-\Omega,-k)] \nonumber \\
&-& E_2^0 [\widehat{e}_2(\Omega,k)+ \widehat{e}^*_2(-\Omega,-k)] \} = 0.
\end{eqnarray}

Similarly, the conjugate of Eq. (\ref{eq9}) yields the correspondent equations for the conjugate of these perturbation fields. If we perform the calculations, we see that the system of four homogeneous equations for $e_1$, $e^*_1$, $e_2$ and $e^*_2$, given by Eqs. (\ref{eq9}), and the corresponding complex conjugates, provide a nontrivial solution if the parameters $k$ and $\Omega$ satisfy the dispersion relation.

\begin{equation}
\label{eq10}
 \left[ \left( k - \frac{\Omega}{\nu_{g1}} \right)^2 - f_1 \right] \left[ \left( k - 
 \frac{\Omega}{\nu_{g2}} \right)^2 - f_2 \right] = C_{XPM},
\end{equation}
with

\begin{eqnarray}
\label{eq11}
f_j = \frac{\beta_j \Omega^2}{2} \left( \frac{\beta_j \Omega^2}{2} + \frac{2\gamma_j  |E_j^0|^2 }{1+\imath \Omega \tau} \right).
\end{eqnarray}
The coupling parameter $C_{XPM}$ is given by

\begin{equation}
\label{eq12}
C_{XPM} = \frac{4\gamma_1\gamma_2\beta_1\beta_2|E^0_1|^2 |E^0_2|^2 \Omega^4}{(1+\imath \Omega \tau)^2}.
\end{equation}

This dispersion relation [Eqs.(\ref{eq10})-(\ref{eq12})] regulates the stability of the steady-state solution with respect to harmonic perturbations. Hence, each MI gain spectrum corresponds to a specific solution of this equation, such that we can qualitatively and quantitatively analyze the effect of the group-velocity mismatch $\delta$ and the delayed nonlinear response time $\tau$ on the instabilities bands. Harmonic perturbations increase exponentially when $k$ acquires an imaginary part once that the gain is defined as $g(\Omega) = 2 \text{ Im}(k)$. A throughly  analysis of the effect of $\delta$ and $\tau$ is performed in the next section.

\section{Discussions and Results}
\label{sec:results}

\begin{figure}[htbp]
\centering
\includegraphics[scale =0.75]{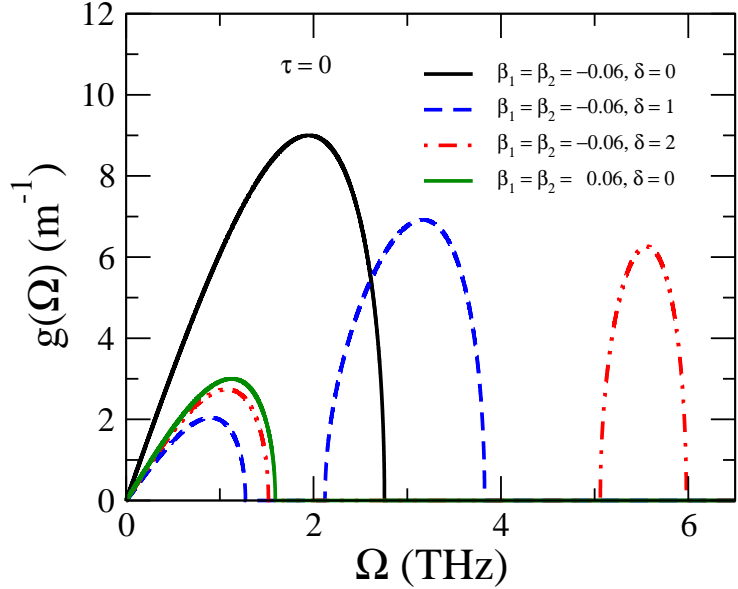}
\caption{(Color online) MI gain spectra in function of the frequencies $\Omega$ (THz) for different values of $\delta$ (ps m$^{-1}$) for $\tau = 0$ ps.}
\label{fig:1}
\end{figure}

For the usual Kerr approach, i.e. $\tau = 0$, we see that the dispersion relation corresponds to a fourth-order polynomial with real coefficients in $k$, once $f_i$ and $C_XPM$ are real numbers, which yields four solutions.  Two of these solutions are necessarily real and, therefore, unimportant to investigations of MI. The two others are possibly a complex conjugate pair of solutions, which would affect the dynamics of the M, generally giving rise to only one unstable gain sideband, once complex solutions appears by pairs and the exponential gain term $e^{g(\Omega)}$ accounts for instability only when the imaginary component of $k$ if positive. Obviously, even for this case, it is not straightforward to obtain analytical relations for the role played by $\delta$ as well as $\tau$ on the XPM-MI. Therefore, we rely on numerical computations.

When not stated otherwise, on the numerical calculations we use the following values for the parameters: the dispersion parameters given by $\beta_1 = \beta_2 = - 0.06 \text{ ps}^2 \text{m}^{-1}$, the nonlinear terms set to $\gamma_1 = \gamma_2 = 0.015 \text{ W}^{-1} \text{m}^{-1}$. The input optical powers as $P_1 = P_2 = 100 \text{ W}$. We fix $\nu_1 = 1\text{ m/ps}$ and vary $\delta$ in the interval [0,10] $\text{ (ps}\text{m}^{-1})$, and the finite nonlinear time delay $\tau$ varying in the range [0,10] $\text{ (ps)}$.  

Aiming to have an overall understanding of the influence of the GVM on the XPM-MI, we plot in Fig. \ref{fig:1} some MI gain spectra for a couple of values of $\delta$. It is possible to check two main effects of the introduction of the GVM term. The first one is that it promotes the appearance of a second instability band at high frequencies values, therefore splitting the single MI band into two instabilities bands when a threshold is attained, as will be clearer later. The second effect if that it reduces the maximum value of the gain. For reference, we also plot the conventional case of both beams experiencing the normal GVD (green solid curve in Fig. \ref{fig:1}). It is possible to check that when $\delta$ increases, the band occurring at lower frequencies evolves to the normal case, slightly increasing its maximum gain as well as the frequency range. One the other hand, the band at higher frequencies reduces their peak and width and move to still higher frequencies. Compare dashed and dotted dashed curves in Fig. \ref{fig:1}.         

\begin{figure}[htbp]
\centering
\fbox{\includegraphics*[scale =0.75]{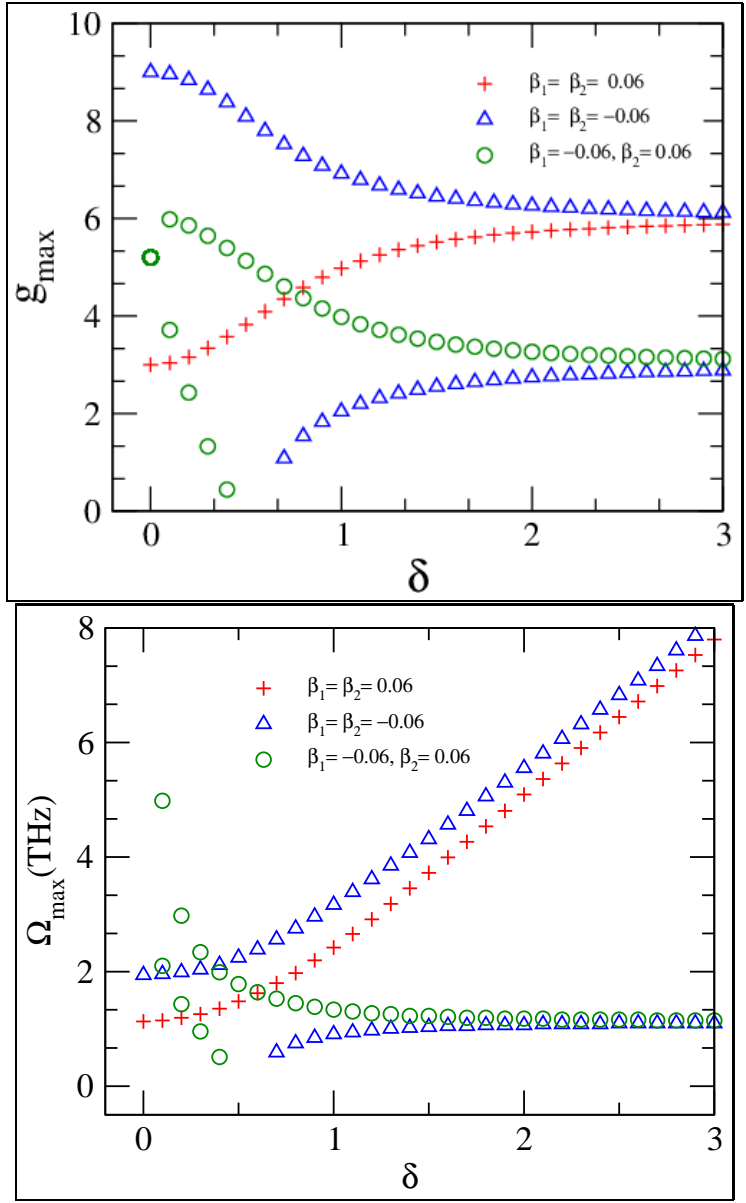}}
\fbox{\includegraphics*[scale =0.75]{fig2b}}
\caption{(Color online) Top: Local maximum for the gain g$_{max}$ (m$^{-1}$) for each instability branch in function of the GVM $\delta$ (ps m$^{-1}$) and conventional Kerr response $\tau = 0$ (ps) for the case of both beams in the normal GVD regime (red pluses), mixed GVD regime (green circles) and both beams in anomalous GVD regime (blue triangles). Bottom: The corresponding frequencies $\Omega_{max}$ (THz) of the maximums values of g$_{max}$.}
\label{fig:2}
\end{figure}

\begin{figure}[htbp]
\centering
\fbox{\includegraphics*[scale =0.75]{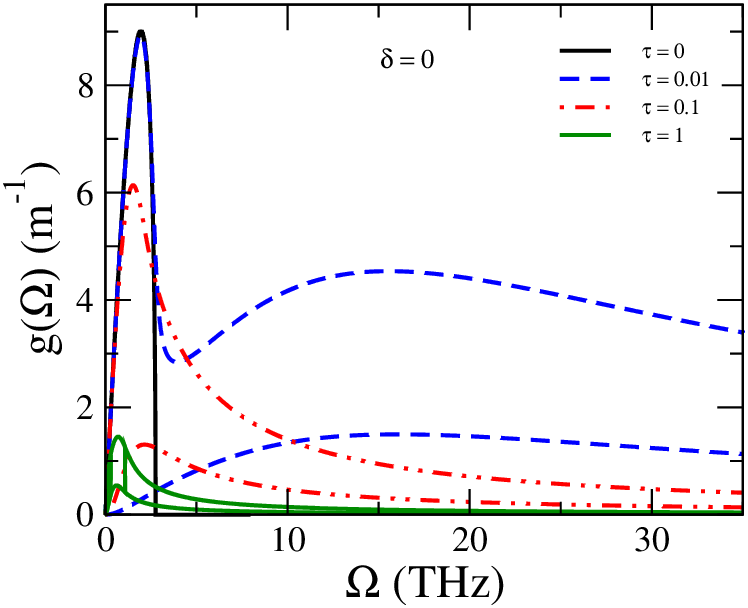}}
\caption{(Color online) MI gain spectra in function of the frequencies $\Omega$ (THz) for different values of $\tau$ (ps) for $\delta = 0$ (ps m$^{-1}$).}
\label{fig:3}
\end{figure}

For a more profound investigation of the solely GVM effect (keeping $\tau = 0$), we plot in Fig. \ref{fig:2} both the peaks (local maximum) of the gain function and their respective frequencies as a function of $\delta$. Here, blue triangles represent the case of both fields experiencing anomalous GVD - which is the topic in this paper, red pluses correspond to the situation where both beams are in the normal GVD and green circles reflect the mixed regime case. We show all of them together for a better distinction of the unique behavior present in the situation investigated in the present work. As stated previously, when we have both beams in anomalous GVD regime, one of the main effect due to $\delta$ is the emergence of a new side band at high frequencies. 

From Fig. \ref{fig:2} (top), this effect is readily noticed once two triangles curves are present for $\delta \gtrapprox 0.75$, given rise to two instabilities bands. We also see that the upper curve decreases with increasing $\delta$, reaching a plateau which corresponds to the normal GVD scenario with the same value of $\delta$ (pluses), whereas the lower triangle curve increases toward another plateau which is the same as the one reached in the case of mixed GVD regimes with the same value for $\delta$, and also to the normal GVD regime with $\delta = 0$, as discussed previously. In this manner, we can state that mainly for the case of both fields experiencing anomalous GVD regime two bands emerge in function of increasing values of $\delta$. One can notice the presence of two bands for the case of mixed GVD for small values of $\delta$ (green circles), but this happens for a tiny range of $\delta$ values and may corresponds just to sporadic peaks and not a proper instability band \cite{canabarro0}.   

\begin{figure}[htbp]
\centering
\fbox{\includegraphics*[scale =0.75]{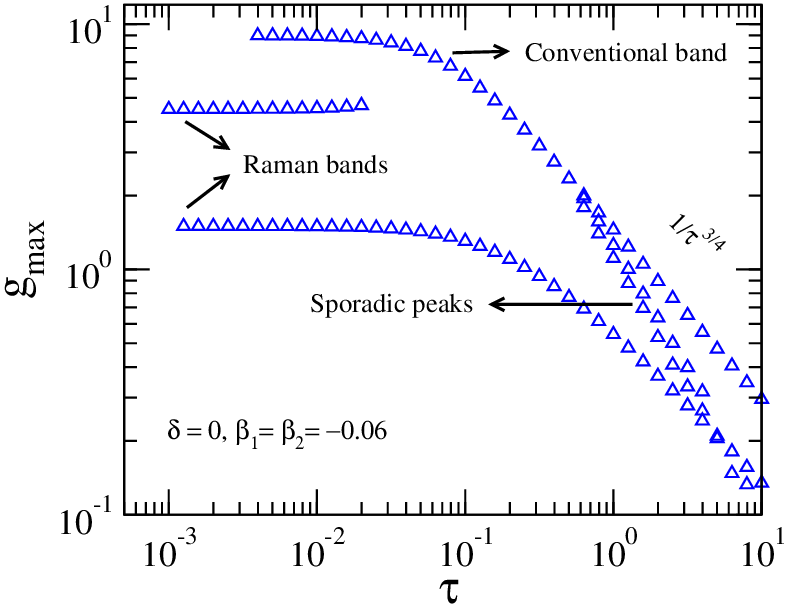}}
\fbox{\includegraphics*[scale =0.75]{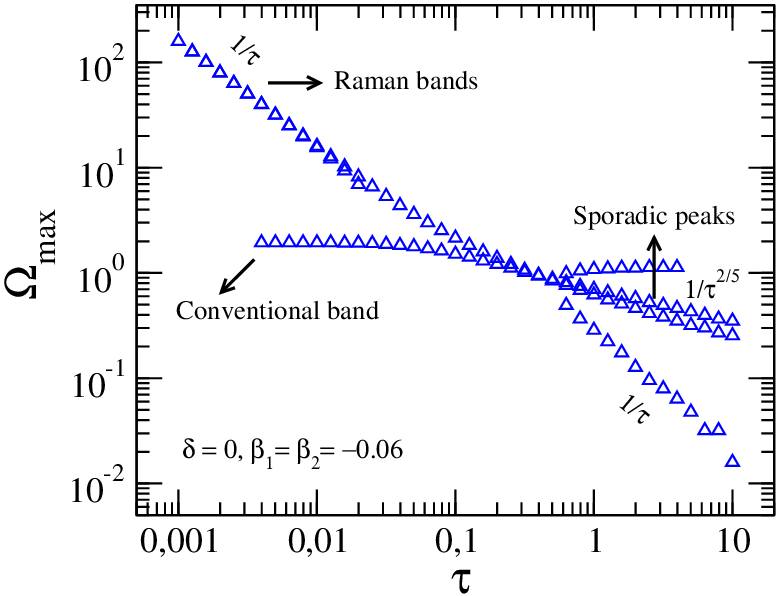}}
\caption{(Color online) Top: Local maximum for the gain g$_{max}$ (m$^{-1}$) for each instability branch in function of $\tau$ for GVM $\delta = 0$ (ps m$^{-1}$) for the case of both beams in the anomalous GVD regime (blue triangles). Bottom: The corresponding frequencies $\Omega_{max}$ (THz) of the maximums values of g$_{max}$.}
\label{fig:4}
\end{figure}

Analyzing Fig. \ref{fig:2} (bottom) we see that the corresponding frequency of the maximum of the instability band occurring at higher frequencies increases linearly with $\delta$, scaling in the same way as in the scenario of both beams in the normal regime (red pluses), whereas the other band practically is not influenced by the GVM in the same way as the case of mixed GVD regimes (green circles). So, the instability band which evolves to the normal case with neglected GVM ($\delta$ = 0) does not change considerably the frequency in which the maximum gain occurs, but the other band moves to higher frequencies at the same pace as $\delta$ is increased.   

\begin{figure}[htbp]
\centering
\fbox{\includegraphics*[scale =0.75]{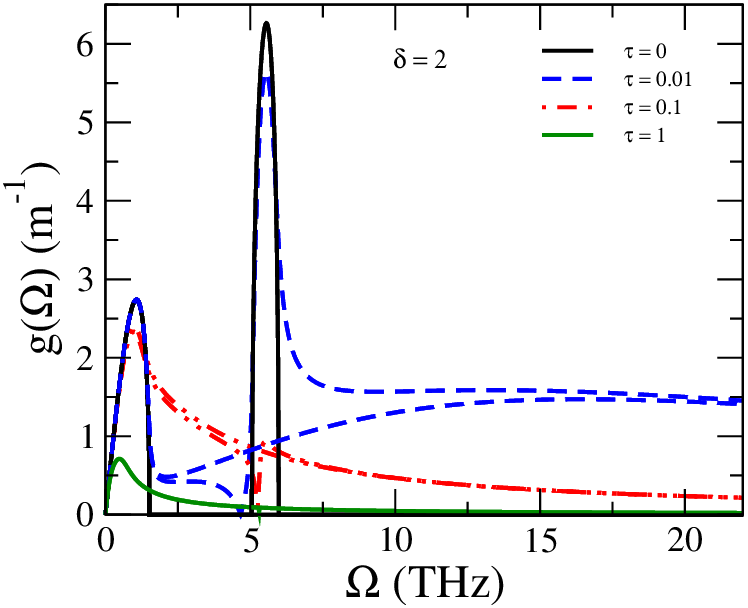}}
\caption{(Color online) MI gain spectra in function of the frequencies $\Omega$ (THz) for different values of $\tau$ (ps) for $\delta = 2$ (ps m$^{-1}$).}
\label{fig:5}
\end{figure}

\begin{figure}[htbp]
\centering
\fbox{\includegraphics*[scale =0.75]{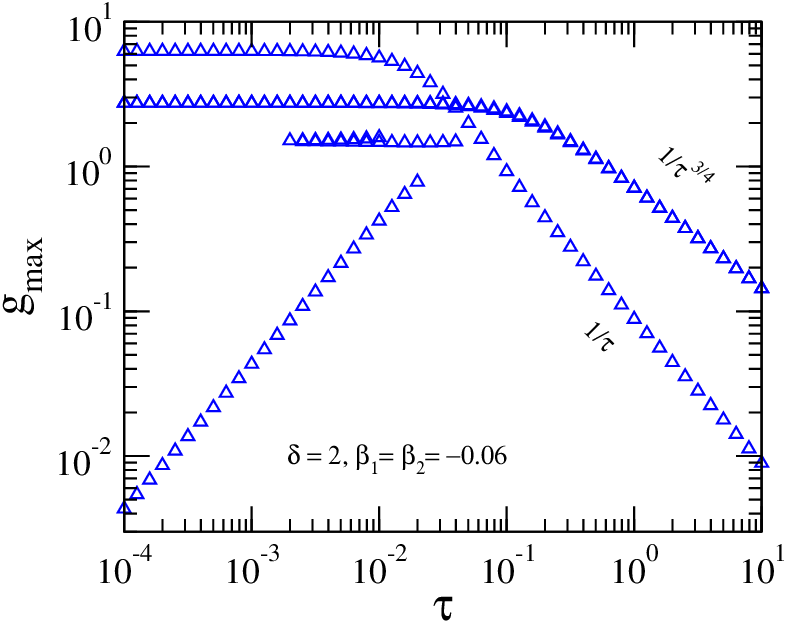}}
\fbox{\includegraphics*[scale =0.75]{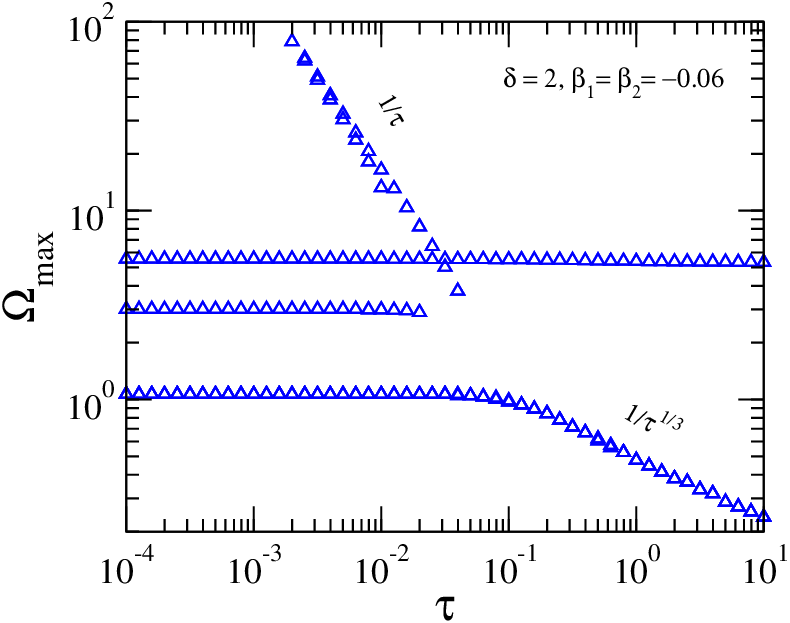}}
\caption{(Color online) Top: Local maximum for the gain g$_{max}$ (m$^{-1}$) for each instability branch in function of $\tau$ for GVM $\delta = 2$ (ps m$^{-1}$) for the case of both beams in anomalous GVD regime (blue triangles). Bottom: The corresponding frequencies $\Omega_{max}$ (THz) of the maximums values of g$_{max}$.}
\label{fig:6}
\end{figure}

Let us pay attention now to the noninstantaneous scenario.  When we add a nonzero response time $(\tau \neq 0)$, the terms $f_i$ [Eq.(\ref{eq11})] and $C_{XPM}$ [Eq.(\ref{eq12})] are now complex numbers, hence the dispersion relation becomes a polynomial equation of degree four with complex coefficients, therefore complex solutions are not obliged to emerge in conjugate pairs anymore. So, it is possible to yield up to four unstable modes (solutions) for a specific frequency $\Omega$ \cite{canabarro0,canabarro2}. 

The general role played by the time delay of the nonlinear response alone (keeping $\delta = 0$) is depicted in Fig. \ref{fig:3}. We notice that the noninstantaneous nonlinear response also split the single MI into more than one instability band. For smaller values of $\tau$ (see dashed blue curves in Fig. \ref{fig:3}), we notice three instabilities bands: one almost the same as the instantaneous case (solid black curve), two occurring at higher frequencies, showing different peaks, but almost centered at the same maximal frequency. Once the origin of the delayed effect is commonly associated with the vibrational Raman effect, we name those new sidebands as Raman bands. 

\begin{figure}[htbp]
\centering
\fbox{\includegraphics*[scale =1.0]{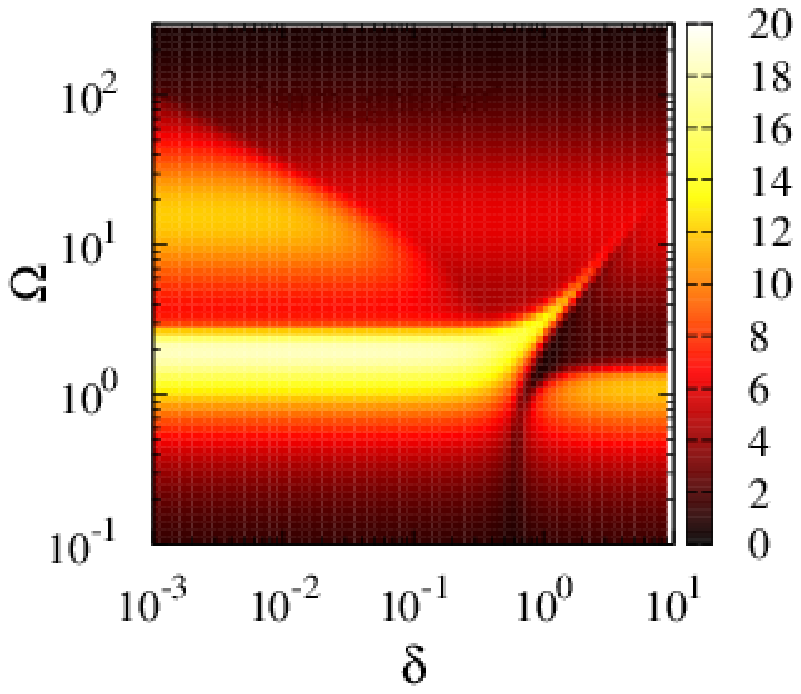}}
\fbox{\includegraphics*[scale =1.0]{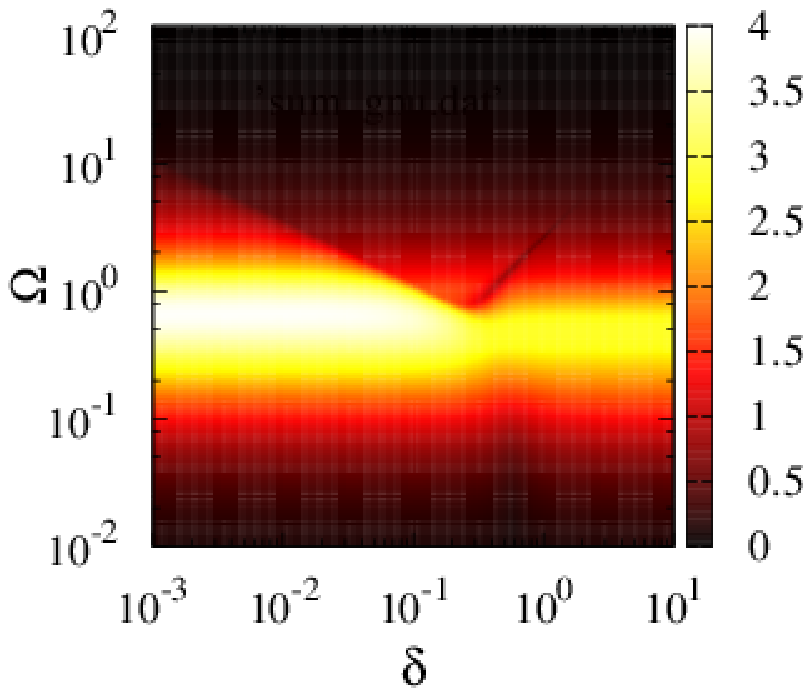}}
\caption{(Color online) Total gain in function of both frequency $\Omega$ and $\delta$ for: Top: $\tau = 0.01$ (ps), Bottom: $\tau = 1$ (ps).}
\label{fig:7}
\end{figure}

As $\tau$ is further increased, one of these new bands tend to merge with the one similar to the instantaneous case, but showing a lower peak, whereas the other also move to lower frequency with an even lower maximum value. In this manner, the main effects of $\tau$ on the XPM-MI can be divided in fast (low values) and slow (high values) regimes \cite{canabarro0,canabarro2}. For $\tau = 1$ (solid green curve) one can see a sporadic peak occurring at low frequency. We shall discuss that in more detail in the sequence.

A more throughly description of the effect of $\tau$ on the modulation instability is done in Fig. \ref{fig:4} as we plot the peaks of the gain function and its respective frequency as a function of $\tau$. From Fig. \ref{fig:4} (top), it is possible to notice the different regimes of fast (low values) and slow (high values) nonlinear response, when a decay in the peaks of gains start appearing. We can use this figure to examine the emergence of more instabilities bands. As we already discussed, small values of $\tau$ give rise to new sidebands. This can be readily noticed for values of $\tau$ around 0.01 as we discussed in the analysis of Fig. \ref{fig:3}. As $\tau$ is further increased, the peaks decrease, scaling according to $1/\tau^{3/4}$. This is a faster decay scaling in comparison to both the mixed GVD regime and the normal GVD case where $g_{max} \propto 1/\tau^{2/3}$ (not shown here, available in \cite{canabarro0} and \cite{canabarro2}, respectively). The sporadic peaks which appear for $\tau \in [1,4]$ decrease in a faster way. As it does not correspond to a proper band, we shall not characterize its behavior any further. 

\begin{figure}[htbp]
\centering
\fbox{\includegraphics*[scale =1.0]{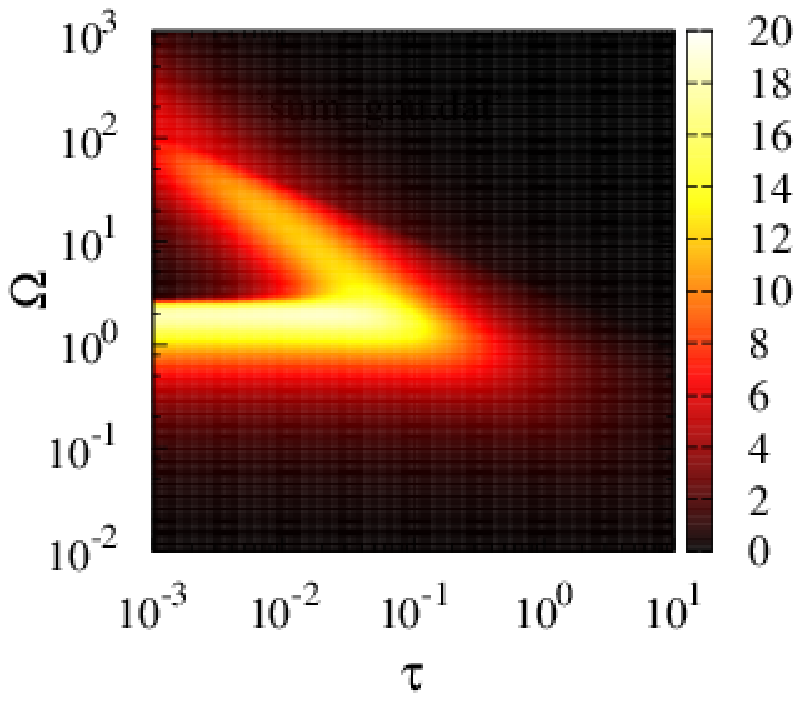}}
\fbox{\includegraphics*[scale =1.0]{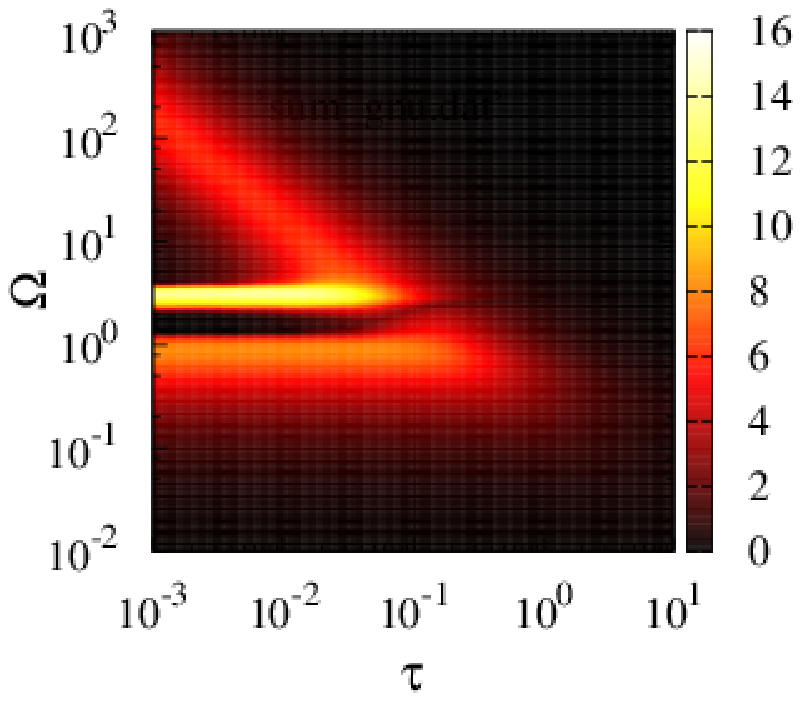}}
\caption{(Color online) Total gain in function of both frequency $\Omega$ and $\tau$ for: Top: $\delta = 0.01$ (ps m$^{-1}$), Bottom: $\delta = 1$ (ps m$^{-1}$).}
\label{fig:8}
\end{figure}

Investigating how the frequencies in which the peaks occur, we see that the band that almost coincides with the conventional band is not dependent of $\tau$ in the fast response regime, see Fig. \ref{fig:4} (bottom), however the frequencies in the Raman bands (which present almost the same frequencies) decay according to $\Omega_{max} \propto 1/\tau$ for the fast regime. For high values of $\tau$ (slow response regime), the merged Raman band keep lowering its central frequency as $1/\tau^{2/5}$, along with the sporadic peaks frequency. The other Raman band keep scaling with the same proportion as in the fast regime, i.e., $\Omega_{max} \propto 1/\tau$. Those dependences within the slow response regime are different for the case of mixed GVD and both beams in normal GVD where one can see a scaling behavior according to $\Omega_{max} \propto 1/\tau^ {1/3}$ \cite{canabarro0,canabarro2}.   


To simultaneously study the effects on the MI, we fixed the walk-off in $\delta = 2$ and plot the gain spectra for some representative values of the delay parameter for fast and slow regimes, as depicted in Fig. \ref{fig:5}. As we have stressed previously for $\delta = 2$ and $\tau = 0$ (Fig. \ref{fig:1},) we observe two instabilities bands, as can be seen again by the solid black curve in Fig. \ref{fig:5}, kept here for a better comparison. For small values of $\tau$, we notice the emergence of two almost identical Raman bands. Between these two peaks, there appears what seems to be another band, but in fact is similar to the sporadic peak that was found previously. See dashed blue curve in Fig. \ref{fig:5}. 


Let us investigate these joint effects.  Fig. \ref{fig:6} (top) shows that in the fast response regime there exist three peaks when $\tau \in [10^{-4}, 10^{-3}]$, the smaller one corresponding to a sporadic peak which increases linearly with $\delta$. For $\tau \in [10^{-3}, 10^{-2}]$, one observes four peaks. Notice that the main bands, which corresponds to the two largest peaks values, are almost independent of $\tau$. When $\tau$ reaches a threshold around $\tau \approx 0.02$, the peaks start decaying and the merged curves scale with $1/\tau$. For mixed GVD regime and both fields in normal GVD regime, this decay occurs as $\tau^{2/3}$ \cite{canabarro0,canabarro2}. The other band peak vanishes in accordance to $1/\tau^ {3/4}$.   
When we investigate as the maximal frequency scales with $\tau$ in Fig. \ref{fig:6} (bottom), we observe that for the fast nonlinear regime the frequencies is not sensibly altered. However, when we are dealing with the slow response regime, the Raman band merged with the high frequency conventional band decays with $1/\tau$, whereas $\Omega_{max} \propto \tau^{1/3}$ for the Raman band which merged with the low frequency conventional band. As we have already stressed so far, this Raman band occurring at high frequency is not present in the other scenarios. However, the last decay is similar to the behaviors demonstrated in \cite{canabarro0,canabarro2}. 

In order to present an heuristic overview of the interplay between the walk-off and the noninstantaneous nonlinear response on the XPM-MI, we present in Figs. \ref{fig:7} and \ref{fig:8} a color plot of the sum of all instabilities bands as function of the frequency and both $\delta$ (Fig. \ref{fig:7}) and $\tau$ (Fig. \ref{fig:8}). 

To have a general view of the role played by the GVM, we plot a color plot of the total gain in function of the frequency and $\delta$ for a small value of the response time $\tau =  0.01$ in Fig. \ref{fig:7} (top). As $\delta$ is increased, we observe that one band (occurring at high frequencies) merges with the band at lower frequencies (the brighter band centered around $\Omega = 2$). When the GVM reaches a value close to $\delta = 1$, we notice that the merged bands move almost linearly with increasing $\delta$ and a remaining band is still present and centered around $\Omega = 1$. For the case of slow responding media ($\tau = 1$) plotted in Fig. \ref{fig:7} (bottom), we can clearly notice that the walk-off effect does not substantially alter the color plot. It slightly narrows the band and reduces its peak. Observe that the brightness is reduced as $\delta$ is increased.       

We performed a similar analysis in Fig. \ref{fig:8}, but this time to study how the gain spectra are influenced by the delay response time $\tau$. In Fig. \ref{fig:8} (top), we use  $\delta = 0.01$ and notice that for fast response (small $\tau$) two distinguishable bands are present, the more intense occurring at lower frequencies. When $\tau$ is further increased, they coalesce in just one band. For $\delta = 1$ the situation is richer, see Fig. \ref{fig:8} (bottom). For small values of $\tau$, we observe three bands. Two of them occurring at high frequencies are the Raman bands and the other is similar to the conventional band. When $\tau$ is increased reaching the slow responding regime, the Raman bands coalesce in just one band. 

Analyzing Fig. \ref{fig:8}, we observe that when $\tau > 1$ the gain is suppressed, which could also be derived from previous analysis. This could be remarkably interesting from a practical point of view once the gain corresponds to the loss of energy and information of the main beam. However, notwithstanding the difficult to manufacture media with high delay time, it also attenuates the main optical field. One of the most promising findings in this study is that these two well distinguishable sidebands may be used as trigger mechanisms for the formation of solitions or even rogue waves.

\section{Conclusions}

In conclusion, we studied modulation instability induced by cross phase modulation in a nonlinear Kerr medium with walk-off and noninstantaneous optical response. We focused on the case in which both copropagating beams are in the anomalous regime of dispersion. In opposition to the cases wherein the optical fields are both either in the normal or mixed regimes, we found that two sidebands types take place for a long range of phase mismatch with a different behavior of the maximum gain. Also, the form in which the maximum gain frequencies of the sidebands vary with respect to the phase mismatch has shown to be remarkably different. Indeed, while some of the bands show a negligible variation as the phase mismatch increases, which is a characteristic in XPM in the mixed regime of dispersion, the other presents an increasing linear behavior, which is normally presented when the pulses are in the normal regime. We also demonstrate that the maximums gains and the corresponding frequencies scale with the finite response time in a different fashion with respect to the scenarios in which both fields experiencing the normal group-velocity dispersion regime and also the case of mixed regimes. Again, two well distinct regimes of fast and slow nonlinear response are shown.

The formation of these two well-defined sideband types in the configuration studied here, instead of only one type, demonstrates that in the anomalous regime the weak perturbation has a higher rate of growth with a more diversified evolution as the waves propagate, when compared to other cases reported in the literature. This fact may favor the appearance of unknown fine localized structures (filamentation), which could also act as a generator of bright solitions. The richer the group of sidebands formed with the present specifications may also represent a more vigorous scenario for the production of rogue waves. In future works we intend to focus on this last point making use of the new findings shown herein.

\section{Acknowledgments}

CNPq (AC's Universal grant No. 423713/2016-7 \& BLB's PQ grant No. 309292/2016-6), 
UFAL (AC's paid license for scientific cooperation at UFRN),
MEC/UFRN (AC's postdoctoral fellowship).

\bibliographystyle{apsrev4-1}
\bibliography{sample}

\begin{thebibliography}{53}%
\makeatletter
\providecommand \@ifxundefined [1]{%
 \@ifx{#1\undefined}
}%
\providecommand \@ifnum [1]{%
 \ifnum #1\expandafter \@firstoftwo
 \else \expandafter \@secondoftwo
 \fi
}%
\providecommand \@ifx [1]{%
 \ifx #1\expandafter \@firstoftwo
 \else \expandafter \@secondoftwo
 \fi
}%
\providecommand \natexlab [1]{#1}%
\providecommand \enquote  [1]{``#1''}%
\providecommand \bibnamefont  [1]{#1}%
\providecommand \bibfnamefont [1]{#1}%
\providecommand \citenamefont [1]{#1}%
\providecommand \href@noop [0]{\@secondoftwo}%
\providecommand \href [0]{\begingroup \@sanitize@url \@href}%
\providecommand \@href[1]{\@@startlink{#1}\@@href}%
\providecommand \@@href[1]{\endgroup#1\@@endlink}%
\providecommand \@sanitize@url [0]{\catcode `\\12\catcode `\$12\catcode
  `\&12\catcode `\#12\catcode `\^12\catcode `\_12\catcode `\%12\relax}%
\providecommand \@@startlink[1]{}%
\providecommand \@@endlink[0]{}%
\providecommand \url  [0]{\begingroup\@sanitize@url \@url }%
\providecommand \@url [1]{\endgroup\@href {#1}{\urlprefix }}%
\providecommand \urlprefix  [0]{URL }%
\providecommand \Eprint [0]{\href }%
\providecommand \doibase [0]{http://dx.doi.org/}%
\providecommand \selectlanguage [0]{\@gobble}%
\providecommand \bibinfo  [0]{\@secondoftwo}%
\providecommand \bibfield  [0]{\@secondoftwo}%
\providecommand \translation [1]{[#1]}%
\providecommand \BibitemOpen [0]{}%
\providecommand \bibitemStop [0]{}%
\providecommand \bibitemNoStop [0]{.\EOS\space}%
\providecommand \EOS [0]{\spacefactor3000\relax}%
\providecommand \BibitemShut  [1]{\csname bibitem#1\endcsname}%
\let\auto@bib@innerbib\@empty
\bibitem [{\citenamefont {Agrawal}(1987)}]{agrawal1}%
  \BibitemOpen
  \bibfield  {author} {\bibinfo {author} {\bibnamefont {Agrawal}, \bibfnamefont
  {G.~P.}},\ }\href {\doibase 10.1103/PhysRevLett.59.880} {\bibfield  {journal}
  {\bibinfo  {journal} {Phys. Rev. Lett.}\ }\textbf {\bibinfo {volume} {59}},\
  \bibinfo {pages} {880} (\bibinfo {year} {1987})}\BibitemShut {NoStop}%
\bibitem [{\citenamefont {Agrawal}(2011)}]{Agrawal:11}%
  \BibitemOpen
  \bibfield  {author} {\bibinfo {author} {\bibnamefont {Agrawal}, \bibfnamefont
  {G.~P.}},\ }\href {\doibase 10.1364/JOSAB.28.0000A1} {\bibfield  {journal}
  {\bibinfo  {journal} {J. Opt. Soc. Am. B}\ }\textbf {\bibinfo {volume}
  {28}},\ \bibinfo {pages} {A1} (\bibinfo {year} {2011})}\BibitemShut {NoStop}%
\bibitem [{\citenamefont {Agrawal}, \citenamefont {Baldeck},\ and\
  \citenamefont {Alfano}(1989)}]{agrawal2}%
  \BibitemOpen
  \bibfield  {author} {\bibinfo {author} {\bibnamefont {Agrawal}, \bibfnamefont
  {G.~P.}}, \bibinfo {author} {\bibnamefont {Baldeck}, \bibfnamefont {P.~L.}},
  \ and\ \bibinfo {author} {\bibnamefont {Alfano}, \bibfnamefont {R.~R.}},\
  }\href {\doibase 10.1103/PhysRevA.39.3406} {\bibfield  {journal} {\bibinfo
  {journal} {Phys. Rev. A}\ }\textbf {\bibinfo {volume} {39}},\ \bibinfo
  {pages} {3406} (\bibinfo {year} {1989})}\BibitemShut {NoStop}%
\bibitem [{\citenamefont {Akhmediev}, \citenamefont {Lederer},\ and\
  \citenamefont {Luther-Davies}(1998)}]{akhmediev98}%
  \BibitemOpen
  \bibfield  {author} {\bibinfo {author} {\bibnamefont {Akhmediev},
  \bibfnamefont {N.~N.}}, \bibinfo {author} {\bibnamefont {Lederer},
  \bibfnamefont {M.~J.}}, \ and\ \bibinfo {author} {\bibnamefont
  {Luther-Davies}, \bibfnamefont {B.}},\ }\href {\doibase
  10.1103/PhysRevE.57.3664} {\bibfield  {journal} {\bibinfo  {journal} {Phys.
  Rev. E}\ }\textbf {\bibinfo {volume} {57}},\ \bibinfo {pages} {3664}
  (\bibinfo {year} {1998})}\BibitemShut {NoStop}%
\bibitem [{\citenamefont {Ali}, \citenamefont {Nithyanandan},\ and\
  \citenamefont {Porsezian}(2015)}]{nithy2}%
  \BibitemOpen
  \bibfield  {author} {\bibinfo {author} {\bibnamefont {Ali}, \bibfnamefont
  {A.~S.}}, \bibinfo {author} {\bibnamefont {Nithyanandan}, \bibfnamefont
  {K.}}, \ and\ \bibinfo {author} {\bibnamefont {Porsezian}, \bibfnamefont
  {K.}},\ }\href {\doibase https://doi.org/10.1016/j.physleta.2014.11.023}
  {\bibfield  {journal} {\bibinfo  {journal} {Physics Letters A}\ }\textbf
  {\bibinfo {volume} {379}},\ \bibinfo {pages} {223 } (\bibinfo {year}
  {2015})}\BibitemShut {NoStop}%
\bibitem [{\citenamefont {Benjamin}\ and\ \citenamefont
  {Feir}(1967)}]{benjamin}%
  \BibitemOpen
  \bibfield  {author} {\bibinfo {author} {\bibnamefont {Benjamin},
  \bibfnamefont {T.~B.}}\ and\ \bibinfo {author} {\bibnamefont {Feir},
  \bibfnamefont {J.~E.}},\ }\href {\doibase 10.1017/S002211206700045X}
  {\bibfield  {journal} {\bibinfo  {journal} {Journal of Fluid Mechanics}\
  }\textbf {\bibinfo {volume} {27}},\ \bibinfo {pages} {417–430} (\bibinfo
  {year} {1967})}\BibitemShut {NoStop}%
\bibitem [{\citenamefont {Biondini}\ and\ \citenamefont
  {Mantzavinos}(2016)}]{biodini}%
  \BibitemOpen
  \bibfield  {author} {\bibinfo {author} {\bibnamefont {Biondini},
  \bibfnamefont {G.}}\ and\ \bibinfo {author} {\bibnamefont {Mantzavinos},
  \bibfnamefont {D.}},\ }\href {\doibase 10.1103/PhysRevLett.116.043902}
  {\bibfield  {journal} {\bibinfo  {journal} {Phys. Rev. Lett.}\ }\textbf
  {\bibinfo {volume} {116}},\ \bibinfo {pages} {043902} (\bibinfo {year}
  {2016})}\BibitemShut {NoStop}%
\bibitem [{\citenamefont {Biswas}\ \emph {et~al.}(2014)\citenamefont {Biswas},
  \citenamefont {Khan}, \citenamefont {Mahmood},\ and\ \citenamefont
  {Belic}}]{BISWAS20143299}%
  \BibitemOpen
  \bibfield  {author} {\bibinfo {author} {\bibnamefont {Biswas}, \bibfnamefont
  {A.}}, \bibinfo {author} {\bibnamefont {Khan}, \bibfnamefont {K.~R.}},
  \bibinfo {author} {\bibnamefont {Mahmood}, \bibfnamefont {M.~F.}}, \ and\
  \bibinfo {author} {\bibnamefont {Belic}, \bibfnamefont {M.}},\ }\href
  {\doibase https://doi.org/10.1016/j.ijleo.2013.12.061} {\bibfield  {journal}
  {\bibinfo  {journal} {Optik - International Journal for Light and Electron
  Optics}\ }\textbf {\bibinfo {volume} {125}},\ \bibinfo {pages} {3299 }
  (\bibinfo {year} {2014})}\BibitemShut {NoStop}%
\bibitem [{\citenamefont {Cambournac}\ \emph {et~al.}(2002)\citenamefont
  {Cambournac}, \citenamefont {Maillotte}, \citenamefont {Lantz}, \citenamefont
  {Dudley},\ and\ \citenamefont {Chauvet}}]{Cambournac02}%
  \BibitemOpen
  \bibfield  {author} {\bibinfo {author} {\bibnamefont {Cambournac},
  \bibfnamefont {C.}}, \bibinfo {author} {\bibnamefont {Maillotte},
  \bibfnamefont {H.}}, \bibinfo {author} {\bibnamefont {Lantz}, \bibfnamefont
  {E.}}, \bibinfo {author} {\bibnamefont {Dudley}, \bibfnamefont {J.~M.}}, \
  and\ \bibinfo {author} {\bibnamefont {Chauvet}, \bibfnamefont {M.}},\ }\href
  {\doibase 10.1364/JOSAB.19.000574} {\bibfield  {journal} {\bibinfo  {journal}
  {J. Opt. Soc. Am. B}\ }\textbf {\bibinfo {volume} {19}},\ \bibinfo {pages}
  {574} (\bibinfo {year} {2002})}\BibitemShut {NoStop}%
\bibitem [{\citenamefont {Canabarro}\ \emph {et~al.}(2016)\citenamefont
  {Canabarro}, \citenamefont {Santos}, \citenamefont {de~Lima~Bernardo},
  \citenamefont {Moura}, \citenamefont {Soares}, \citenamefont {de~Lima},
  \citenamefont {Gl\'eria},\ and\ \citenamefont {Lyra}}]{canabarro0}%
  \BibitemOpen
  \bibfield  {author} {\bibinfo {author} {\bibnamefont {Canabarro},
  \bibfnamefont {A.}}, \bibinfo {author} {\bibnamefont {Santos}, \bibfnamefont
  {B.}}, \bibinfo {author} {\bibnamefont {de~Lima~Bernardo}, \bibfnamefont
  {B.}}, \bibinfo {author} {\bibnamefont {Moura}, \bibfnamefont {A.~L.}},
  \bibinfo {author} {\bibnamefont {Soares}, \bibfnamefont {W.~C.}}, \bibinfo
  {author} {\bibnamefont {de~Lima}, \bibfnamefont {E.}}, \bibinfo {author}
  {\bibnamefont {Gl\'eria}, \bibfnamefont {I.}}, \ and\ \bibinfo {author}
  {\bibnamefont {Lyra}, \bibfnamefont {M.~L.}},\ }\href {\doibase
  10.1103/PhysRevA.93.023834} {\bibfield  {journal} {\bibinfo  {journal} {Phys.
  Rev. A}\ }\textbf {\bibinfo {volume} {93}},\ \bibinfo {pages} {023834}
  (\bibinfo {year} {2016})}\BibitemShut {NoStop}%
\bibitem [{\citenamefont {Canabarro}\ \emph {et~al.}(2010)\citenamefont
  {Canabarro}, \citenamefont {Santos}, \citenamefont {Gleria}, \citenamefont
  {Lyra},\ and\ \citenamefont {Sombra}}]{canabarro2}%
  \BibitemOpen
  \bibfield  {author} {\bibinfo {author} {\bibnamefont {Canabarro},
  \bibfnamefont {A.~A.}}, \bibinfo {author} {\bibnamefont {Santos},
  \bibfnamefont {B.}}, \bibinfo {author} {\bibnamefont {Gleria}, \bibfnamefont
  {I.}}, \bibinfo {author} {\bibnamefont {Lyra}, \bibfnamefont {M.~L.}}, \ and\
  \bibinfo {author} {\bibnamefont {Sombra}, \bibfnamefont {A.~S.~B.}},\ }\href
  {\doibase 10.1364/JOSAB.27.001878} {\bibfield  {journal} {\bibinfo  {journal}
  {J. Opt. Soc. Am. B}\ }\textbf {\bibinfo {volume} {27}},\ \bibinfo {pages}
  {1878} (\bibinfo {year} {2010})}\BibitemShut {NoStop}%
\bibitem [{\citenamefont {Carbou}\ and\ \citenamefont
  {Hanouzet}(2006)}]{carbou2006}%
  \BibitemOpen
  \bibfield  {author} {\bibinfo {author} {\bibnamefont {Carbou}, \bibfnamefont
  {G.}}\ and\ \bibinfo {author} {\bibnamefont {Hanouzet}, \bibfnamefont {B.}},\
  }\href {https://projecteuclid.org:443/euclid.cms/1154635527} {\bibfield
  {journal} {\bibinfo  {journal} {Commun. Math. Sci.}\ }\textbf {\bibinfo
  {volume} {4}},\ \bibinfo {pages} {331} (\bibinfo {year} {2006})}\BibitemShut
  {NoStop}%
\bibitem [{\citenamefont {Stepi\ifmmode~\acute{c}\else \'{c}\fi{}}\ \emph
  {et~al.}(2008)\citenamefont {Stepi\ifmmode~\acute{c}\else \'{c}\fi{}},
  \citenamefont {Maluckov}, \citenamefont {Stojanovi\ifmmode~\acute{c}\else
  \'{c}\fi{}}, \citenamefont {Chen},\ and\ \citenamefont {Kip}}]{stepiPRA}%
  \BibitemOpen
  \bibfield  {author} {\bibinfo {author} {\bibnamefont
  {Stepi\ifmmode~\acute{c}\else \'{c}\fi{}}, \bibfnamefont {M.}}, \bibinfo
  {author} {\bibnamefont {Maluckov}, \bibfnamefont {A.}}, \bibinfo {author}
  {\bibnamefont {Stojanovi\ifmmode~\acute{c}\else \'{c}\fi{}}, \bibfnamefont
  {M.}}, \bibinfo {author} {\bibnamefont {Chen}, \bibfnamefont {F.}}, \ and\
  \bibinfo {author} {\bibnamefont {Kip}, \bibfnamefont {D.}},\ }\href {\doibase
  10.1103/PhysRevA.78.043819} {\bibfield  {journal} {\bibinfo  {journal} {Phys.
  Rev. A}\ }\textbf {\bibinfo {volume} {78}},\ \bibinfo {pages} {043819}
  (\bibinfo {year} {2008})}\BibitemShut {NoStop}%
\bibitem [{\citenamefont {Chou}\ and\ \citenamefont {Shih}(2009)}]{chou09}%
  \BibitemOpen
  \bibfield  {author} {\bibinfo {author} {\bibnamefont {Chou}, \bibfnamefont
  {C.-S.}}\ and\ \bibinfo {author} {\bibnamefont {Shih}, \bibfnamefont
  {M.-F.}},\ }\href {http://stacks.iop.org/1464-4258/11/i=10/a=105204}
  {\bibfield  {journal} {\bibinfo  {journal} {Journal of Optics A: Pure and
  Applied Optics}\ }\textbf {\bibinfo {volume} {11}},\ \bibinfo {pages}
  {105204} (\bibinfo {year} {2009})}\BibitemShut {NoStop}%
\bibitem [{\citenamefont {Chow}, \citenamefont {Wong},\ and\ \citenamefont
  {Lam}(2008)}]{chow1}%
  \BibitemOpen
  \bibfield  {author} {\bibinfo {author} {\bibnamefont {Chow}, \bibfnamefont
  {K.}}, \bibinfo {author} {\bibnamefont {Wong}, \bibfnamefont {K.}}, \ and\
  \bibinfo {author} {\bibnamefont {Lam}, \bibfnamefont {K.}},\ }\href {\doibase
  https://doi.org/10.1016/j.physleta.2008.04.057} {\bibfield  {journal}
  {\bibinfo  {journal} {Physics Letters A}\ }\textbf {\bibinfo {volume}
  {372}},\ \bibinfo {pages} {4596 } (\bibinfo {year} {2008})}\BibitemShut
  {NoStop}%
\bibitem [{\citenamefont {Conti}\ \emph {et~al.}(2010)\citenamefont {Conti},
  \citenamefont {Schmidt}, \citenamefont {Russell},\ and\ \citenamefont
  {Biancalana}}]{contiPRL}%
  \BibitemOpen
  \bibfield  {author} {\bibinfo {author} {\bibnamefont {Conti}, \bibfnamefont
  {C.}}, \bibinfo {author} {\bibnamefont {Schmidt}, \bibfnamefont {M.~A.}},
  \bibinfo {author} {\bibnamefont {Russell}, \bibfnamefont {P.~S.~J.}}, \ and\
  \bibinfo {author} {\bibnamefont {Biancalana}, \bibfnamefont {F.}},\ }\href
  {\doibase 10.1103/PhysRevLett.105.263902} {\bibfield  {journal} {\bibinfo
  {journal} {Phys. Rev. Lett.}\ }\textbf {\bibinfo {volume} {105}},\ \bibinfo
  {pages} {263902} (\bibinfo {year} {2010})}\BibitemShut {NoStop}%
\bibitem [{\citenamefont {Dysthe}, \citenamefont {Krogstad},\ and\
  \citenamefont {MÃ¼ller}(2008)}]{dysthe}%
  \BibitemOpen
  \bibfield  {author} {\bibinfo {author} {\bibnamefont {Dysthe}, \bibfnamefont
  {K.}}, \bibinfo {author} {\bibnamefont {Krogstad}, \bibfnamefont {H.~E.}}, \
  and\ \bibinfo {author} {\bibnamefont {MÃ¼ller}, \bibfnamefont {P.}},\
  }\href {\doibase 10.1146/annurev.fluid.40.111406.102203} {\bibfield
  {journal} {\bibinfo  {journal} {Annual Review of Fluid Mechanics}\ }\textbf
  {\bibinfo {volume} {40}},\ \bibinfo {pages} {287} (\bibinfo {year} {2008})},\
  \Eprint
  {http://arxiv.org/abs/https://doi.org/10.1146/annurev.fluid.40.111406.102203}
  {https://doi.org/10.1146/annurev.fluid.40.111406.102203} \BibitemShut
  {NoStop}%
\bibitem [{\citenamefont {Everitt}\ \emph {et~al.}(2017)\citenamefont
  {Everitt}, \citenamefont {Sooriyabandara}, \citenamefont {Guasoni},
  \citenamefont {Wigley}, \citenamefont {Wei}, \citenamefont {McDonald},
  \citenamefont {Hardman}, \citenamefont {Manju}, \citenamefont {Close},
  \citenamefont {Kuhn}, \citenamefont {Szigeti}, \citenamefont {Kivshar},\ and\
  \citenamefont {Robins}}]{everitt}%
  \BibitemOpen
  \bibfield  {author} {\bibinfo {author} {\bibnamefont {Everitt}, \bibfnamefont
  {P.~J.}}, \bibinfo {author} {\bibnamefont {Sooriyabandara}, \bibfnamefont
  {M.~A.}}, \bibinfo {author} {\bibnamefont {Guasoni}, \bibfnamefont {M.}},
  \bibinfo {author} {\bibnamefont {Wigley}, \bibfnamefont {P.~B.}}, \bibinfo
  {author} {\bibnamefont {Wei}, \bibfnamefont {C.~H.}}, \bibinfo {author}
  {\bibnamefont {McDonald}, \bibfnamefont {G.~D.}}, \bibinfo {author}
  {\bibnamefont {Hardman}, \bibfnamefont {K.~S.}}, \bibinfo {author}
  {\bibnamefont {Manju}, \bibfnamefont {P.}}, \bibinfo {author} {\bibnamefont
  {Close}, \bibfnamefont {J.~D.}}, \bibinfo {author} {\bibnamefont {Kuhn},
  \bibfnamefont {C.~C.~N.}}, \bibinfo {author} {\bibnamefont {Szigeti},
  \bibfnamefont {S.~S.}}, \bibinfo {author} {\bibnamefont {Kivshar},
  \bibfnamefont {Y.~S.}}, \ and\ \bibinfo {author} {\bibnamefont {Robins},
  \bibfnamefont {N.~P.}},\ }\href {\doibase 10.1103/PhysRevA.96.041601}
  {\bibfield  {journal} {\bibinfo  {journal} {Phys. Rev. A}\ }\textbf {\bibinfo
  {volume} {96}},\ \bibinfo {pages} {041601} (\bibinfo {year}
  {2017})}\BibitemShut {NoStop}%
\bibitem [{\citenamefont {Fushman}\ and\ \citenamefont
  {Vu{\v{c}}kovi{\'{c}}}(2007)}]{Fushman07}%
  \BibitemOpen
  \bibfield  {author} {\bibinfo {author} {\bibnamefont {Fushman}, \bibfnamefont
  {I.}}\ and\ \bibinfo {author} {\bibnamefont {Vu{\v{c}}kovi{\'{c}}},
  \bibfnamefont {J.}},\ }\href {\doibase 10.1364/OE.15.005559} {\bibfield
  {journal} {\bibinfo  {journal} {Opt. Express}\ }\textbf {\bibinfo {volume}
  {15}},\ \bibinfo {pages} {5559} (\bibinfo {year} {2007})}\BibitemShut
  {NoStop}%
\bibitem [{\citenamefont {Hallaji}\ \emph {et~al.}(2015)\citenamefont
  {Hallaji}, \citenamefont {Feizpour}, \citenamefont {Dmochowski},
  \citenamefont {Sinclair},\ and\ \citenamefont {Steinberg}}]{Hallaji15}%
  \BibitemOpen
  \bibfield  {author} {\bibinfo {author} {\bibnamefont {Hallaji}, \bibfnamefont
  {M.}}, \bibinfo {author} {\bibnamefont {Feizpour}, \bibfnamefont {A.}},
  \bibinfo {author} {\bibnamefont {Dmochowski}, \bibfnamefont {G.}}, \bibinfo
  {author} {\bibnamefont {Sinclair}, \bibfnamefont {J.}}, \ and\ \bibinfo
  {author} {\bibnamefont {Steinberg}, \bibfnamefont {A.~M.}},\ }in\ \href
  {http://www.osapublishing.org/abstract.cfm?URI=CLEO_Europe-2015-CD_9_2}
  {\emph {\bibinfo {booktitle} {2015 European Conference on Lasers and
  Electro-Optics - European Quantum Electronics Conference}}}\ (\bibinfo
  {publisher} {Optical Society of America},\ \bibinfo {year}
  {2015})\BibitemShut {NoStop}%
\bibitem [{\citenamefont {Harris}\ and\ \citenamefont
  {Yamamoto}(1998)}]{harris}%
  \BibitemOpen
  \bibfield  {author} {\bibinfo {author} {\bibnamefont {Harris}, \bibfnamefont
  {S.~E.}}\ and\ \bibinfo {author} {\bibnamefont {Yamamoto}, \bibfnamefont
  {Y.}},\ }\href {\doibase 10.1103/PhysRevLett.81.3611} {\bibfield  {journal}
  {\bibinfo  {journal} {Phys. Rev. Lett.}\ }\textbf {\bibinfo {volume} {81}},\
  \bibinfo {pages} {3611} (\bibinfo {year} {1998})}\BibitemShut {NoStop}%
\bibitem [{\citenamefont {Janssen}(2003)}]{Janssen}%
  \BibitemOpen
  \bibfield  {author} {\bibinfo {author} {\bibnamefont {Janssen}, \bibfnamefont
  {P.~A. E.~M.}},\ }\href {\doibase
  10.1175/1520-0485(2003)33<863:NFIAFW>2.0.CO;2} {\bibfield  {journal}
  {\bibinfo  {journal} {Journal of Physical Oceanography}\ }\textbf {\bibinfo
  {volume} {33}},\ \bibinfo {pages} {863} (\bibinfo {year} {2003})},\ \Eprint
  {http://arxiv.org/abs/https://doi.org/10.1175/1520-0485(2003)33<863:NFIAFW>2.0.CO;2}
  {https://doi.org/10.1175/1520-0485(2003)33<863:NFIAFW>2.0.CO;2} \BibitemShut
  {NoStop}%
\bibitem [{\citenamefont {Kengne}, \citenamefont {Chui},\ and\ \citenamefont
  {Liu}(2006)}]{PhysRevE.74.036614}%
  \BibitemOpen
  \bibfield  {author} {\bibinfo {author} {\bibnamefont {Kengne}, \bibfnamefont
  {E.}}, \bibinfo {author} {\bibnamefont {Chui}, \bibfnamefont {S.~T.}}, \ and\
  \bibinfo {author} {\bibnamefont {Liu}, \bibfnamefont {W.~M.}},\ }\href
  {\doibase 10.1103/PhysRevE.74.036614} {\bibfield  {journal} {\bibinfo
  {journal} {Phys. Rev. E}\ }\textbf {\bibinfo {volume} {74}},\ \bibinfo
  {pages} {036614} (\bibinfo {year} {2006})}\BibitemShut {NoStop}%
\bibitem [{\citenamefont {Kivshar}\ and\ \citenamefont {Agrawal}(2003)}]{kiv}%
  \BibitemOpen
  \bibfield  {author} {\bibinfo {author} {\bibnamefont {Kivshar}, \bibfnamefont
  {Y.}}\ and\ \bibinfo {author} {\bibnamefont {Agrawal}, \bibfnamefont {G.}},\
  }\href {https://books.google.com.br/books?id=zzWgibj4ypsC} {\emph {\bibinfo
  {title} {Optical Solitons: From Fibers to Photonic Crystals}}}\ (\bibinfo
  {publisher} {Elsevier Science},\ \bibinfo {year} {2003})\BibitemShut
  {NoStop}%
\bibitem [{\citenamefont {Kr{\'{o}}likowski}\ \emph {et~al.}(2004)\citenamefont
  {Kr{\'{o}}likowski}, \citenamefont {Bang}, \citenamefont {Nikolov},
  \citenamefont {Neshev}, \citenamefont {Wyller}, \citenamefont {Rasmussen},\
  and\ \citenamefont {Edmundson}}]{edmunson04}%
  \BibitemOpen
  \bibfield  {author} {\bibinfo {author} {\bibnamefont {Kr{\'{o}}likowski},
  \bibfnamefont {W.}}, \bibinfo {author} {\bibnamefont {Bang}, \bibfnamefont
  {O.}}, \bibinfo {author} {\bibnamefont {Nikolov}, \bibfnamefont {N.~I.}},
  \bibinfo {author} {\bibnamefont {Neshev}, \bibfnamefont {D.}}, \bibinfo
  {author} {\bibnamefont {Wyller}, \bibfnamefont {J.}}, \bibinfo {author}
  {\bibnamefont {Rasmussen}, \bibfnamefont {J.~J.}}, \ and\ \bibinfo {author}
  {\bibnamefont {Edmundson}, \bibfnamefont {D.}},\ }\href
  {http://stacks.iop.org/1464-4266/6/i=5/a=017} {\bibfield  {journal} {\bibinfo
   {journal} {Journal of Optics B: Quantum and Semiclassical Optics}\ }\textbf
  {\bibinfo {volume} {6}},\ \bibinfo {pages} {S288} (\bibinfo {year}
  {2004})}\BibitemShut {NoStop}%
\bibitem [{\citenamefont {Lin}\ and\ \citenamefont {Agrawal}(2004)}]{lin04}%
  \BibitemOpen
  \bibfield  {author} {\bibinfo {author} {\bibnamefont {Lin}, \bibfnamefont
  {Q.}}\ and\ \bibinfo {author} {\bibnamefont {Agrawal}, \bibfnamefont
  {G.~P.}},\ }\href {\doibase 10.1109/JQE.2004.830198} {\bibfield  {journal}
  {\bibinfo  {journal} {IEEE Journal of Quantum Electronics}\ }\textbf
  {\bibinfo {volume} {40}},\ \bibinfo {pages} {958} (\bibinfo {year}
  {2004})}\BibitemShut {NoStop}%
\bibitem [{\citenamefont {Litvak}(1969)}]{litvak}%
  \BibitemOpen
  \bibfield  {author} {\bibinfo {author} {\bibnamefont {Litvak}, \bibfnamefont
  {A.~G.}},\ }\href@noop {} {\bibfield  {journal} {\bibinfo  {journal} {Zh.
  Eksp. Teor. Fiz.}\ }\textbf {\bibinfo {volume} {57}},\ \bibinfo {pages} {629}
  (\bibinfo {year} {1969})}\BibitemShut {NoStop}%
\bibitem [{\citenamefont {Liu}, \citenamefont {Haus},\ and\ \citenamefont
  {Shahriar}(2008)}]{LIU20082907}%
  \BibitemOpen
  \bibfield  {author} {\bibinfo {author} {\bibnamefont {Liu}, \bibfnamefont
  {X.}}, \bibinfo {author} {\bibnamefont {Haus}, \bibfnamefont {J.~W.}}, \ and\
  \bibinfo {author} {\bibnamefont {Shahriar}, \bibfnamefont {S.~M.}},\ }\href
  {\doibase https://doi.org/10.1016/j.optcom.2008.01.026} {\bibfield  {journal}
  {\bibinfo  {journal} {Optics Communications}\ }\textbf {\bibinfo {volume}
  {281}},\ \bibinfo {pages} {2907} (\bibinfo {year} {2008})}\BibitemShut
  {NoStop}%
\bibitem [{\citenamefont {Mollenauer}, \citenamefont {Stolen},\ and\
  \citenamefont {Gordon}(1980)}]{moll}%
  \BibitemOpen
  \bibfield  {author} {\bibinfo {author} {\bibnamefont {Mollenauer},
  \bibfnamefont {L.~F.}}, \bibinfo {author} {\bibnamefont {Stolen},
  \bibfnamefont {R.~H.}}, \ and\ \bibinfo {author} {\bibnamefont {Gordon},
  \bibfnamefont {J.~P.}},\ }\href {\doibase 10.1103/PhysRevLett.45.1095}
  {\bibfield  {journal} {\bibinfo  {journal} {Phys. Rev. Lett.}\ }\textbf
  {\bibinfo {volume} {45}},\ \bibinfo {pages} {1095} (\bibinfo {year}
  {1980})}\BibitemShut {NoStop}%
\bibitem [{\citenamefont {Nithyanandan}\ and\ \citenamefont
  {Porsezian}(2013)}]{NITHY3}%
  \BibitemOpen
  \bibfield  {author} {\bibinfo {author} {\bibnamefont {Nithyanandan},
  \bibfnamefont {K.}}\ and\ \bibinfo {author} {\bibnamefont {Porsezian},
  \bibfnamefont {K.}},\ }\href {\doibase
  https://doi.org/10.1016/j.optcom.2013.04.018} {\bibfield  {journal} {\bibinfo
   {journal} {Optics Communications}\ }\textbf {\bibinfo {volume} {303}},\
  \bibinfo {pages} {46} (\bibinfo {year} {2013})}\BibitemShut {NoStop}%
\bibitem [{\citenamefont {Nithyanandan}\ \emph {et~al.}(2012)\citenamefont
  {Nithyanandan}, \citenamefont {Raja}, \citenamefont {Porsezian},\ and\
  \citenamefont {Kalithasan}}]{nithy1}%
  \BibitemOpen
  \bibfield  {author} {\bibinfo {author} {\bibnamefont {Nithyanandan},
  \bibfnamefont {K.}}, \bibinfo {author} {\bibnamefont {Raja}, \bibfnamefont
  {R.~V.~J.}}, \bibinfo {author} {\bibnamefont {Porsezian}, \bibfnamefont
  {K.}}, \ and\ \bibinfo {author} {\bibnamefont {Kalithasan}, \bibfnamefont
  {B.}},\ }\href {\doibase 10.1103/PhysRevA.86.023827} {\bibfield  {journal}
  {\bibinfo  {journal} {Phys. Rev. A}\ }\textbf {\bibinfo {volume} {86}},\
  \bibinfo {pages} {023827} (\bibinfo {year} {2012})}\BibitemShut {NoStop}%
\bibitem [{\citenamefont {Onorato}, \citenamefont {Osborne},\ and\
  \citenamefont {Serio}(2006)}]{onoratoPRL}%
  \BibitemOpen
  \bibfield  {author} {\bibinfo {author} {\bibnamefont {Onorato}, \bibfnamefont
  {M.}}, \bibinfo {author} {\bibnamefont {Osborne}, \bibfnamefont {A.~R.}}, \
  and\ \bibinfo {author} {\bibnamefont {Serio}, \bibfnamefont {M.}},\ }\href
  {\doibase 10.1103/PhysRevLett.96.014503} {\bibfield  {journal} {\bibinfo
  {journal} {Phys. Rev. Lett.}\ }\textbf {\bibinfo {volume} {96}},\ \bibinfo
  {pages} {014503} (\bibinfo {year} {2006})}\BibitemShut {NoStop}%
\bibitem [{\citenamefont {Onorato}\ \emph {et~al.}(2001)\citenamefont
  {Onorato}, \citenamefont {Osborne}, \citenamefont {Serio},\ and\
  \citenamefont {Bertone}}]{onoratoPRL2}%
  \BibitemOpen
  \bibfield  {author} {\bibinfo {author} {\bibnamefont {Onorato}, \bibfnamefont
  {M.}}, \bibinfo {author} {\bibnamefont {Osborne}, \bibfnamefont {A.~R.}},
  \bibinfo {author} {\bibnamefont {Serio}, \bibfnamefont {M.}}, \ and\ \bibinfo
  {author} {\bibnamefont {Bertone}, \bibfnamefont {S.}},\ }\href {\doibase
  10.1103/PhysRevLett.86.5831} {\bibfield  {journal} {\bibinfo  {journal}
  {Phys. Rev. Lett.}\ }\textbf {\bibinfo {volume} {86}},\ \bibinfo {pages}
  {5831} (\bibinfo {year} {2001})}\BibitemShut {NoStop}%
\bibitem [{\citenamefont {Potasek}(1987)}]{Potasek87}%
  \BibitemOpen
  \bibfield  {author} {\bibinfo {author} {\bibnamefont {Potasek}, \bibfnamefont
  {M.~J.}},\ }\href {\doibase 10.1364/OL.12.000921} {\bibfield  {journal}
  {\bibinfo  {journal} {Opt. Lett.}\ }\textbf {\bibinfo {volume} {12}},\
  \bibinfo {pages} {921} (\bibinfo {year} {1987})}\BibitemShut {NoStop}%
\bibitem [{\citenamefont {Priya}\ and\ \citenamefont
  {Senthilvelan}(2015)}]{priya}%
  \BibitemOpen
  \bibfield  {author} {\bibinfo {author} {\bibnamefont {Priya}, \bibfnamefont
  {N.~V.}}\ and\ \bibinfo {author} {\bibnamefont {Senthilvelan}, \bibfnamefont
  {M.}},\ }\href {\doibase https://doi.org/10.1016/j.wavemoti.2014.12.001}
  {\bibfield  {journal} {\bibinfo  {journal} {Wave Motion}\ }\textbf {\bibinfo
  {volume} {54}},\ \bibinfo {pages} {125 } (\bibinfo {year}
  {2015})}\BibitemShut {NoStop}%
\bibitem [{\citenamefont {Raja}\ \emph {et~al.}(2010)\citenamefont {Raja},
  \citenamefont {Husakou}, \citenamefont {Hermann},\ and\ \citenamefont
  {Porsezian}}]{Vasantha10}%
  \BibitemOpen
  \bibfield  {author} {\bibinfo {author} {\bibnamefont {Raja}, \bibfnamefont
  {R.~V.~J.}}, \bibinfo {author} {\bibnamefont {Husakou}, \bibfnamefont {A.}},
  \bibinfo {author} {\bibnamefont {Hermann}, \bibfnamefont {J.}}, \ and\
  \bibinfo {author} {\bibnamefont {Porsezian}, \bibfnamefont {K.}},\ }\href
  {\doibase 10.1364/JOSAB.27.001763} {\bibfield  {journal} {\bibinfo  {journal}
  {J. Opt. Soc. Am. B}\ }\textbf {\bibinfo {volume} {27}},\ \bibinfo {pages}
  {1763} (\bibinfo {year} {2010})}\BibitemShut {NoStop}%
\bibitem [{\citenamefont {Rothenberg}(1990)}]{rothen}%
  \BibitemOpen
  \bibfield  {author} {\bibinfo {author} {\bibnamefont {Rothenberg},
  \bibfnamefont {J.~E.}},\ }\href {\doibase 10.1103/PhysRevA.42.682} {\bibfield
   {journal} {\bibinfo  {journal} {Phys. Rev. A}\ }\textbf {\bibinfo {volume}
  {42}},\ \bibinfo {pages} {682} (\bibinfo {year} {1990})}\BibitemShut
  {NoStop}%
\bibitem [{\citenamefont {Schmidt}\ and\ \citenamefont
  {Imamoglu}(1996)}]{Schmidt96}%
  \BibitemOpen
  \bibfield  {author} {\bibinfo {author} {\bibnamefont {Schmidt}, \bibfnamefont
  {H.}}\ and\ \bibinfo {author} {\bibnamefont {Imamoglu}, \bibfnamefont {A.}},\
  }\href {\doibase 10.1364/OL.21.001936} {\bibfield  {journal} {\bibinfo
  {journal} {Opt. Lett.}\ }\textbf {\bibinfo {volume} {21}},\ \bibinfo {pages}
  {1936} (\bibinfo {year} {1996})}\BibitemShut {NoStop}%
\bibitem [{\citenamefont {da~Silva}, \citenamefont {Canabarro},\ and\
  \citenamefont {de~Lima~Bernardo}(2015)}]{canabarro1}%
  \BibitemOpen
  \bibfield  {author} {\bibinfo {author} {\bibnamefont {da~Silva},
  \bibfnamefont {G.}}, \bibinfo {author} {\bibnamefont {Canabarro},
  \bibfnamefont {A.}}, \ and\ \bibinfo {author} {\bibnamefont
  {de~Lima~Bernardo}, \bibfnamefont {B.}},\ }\href {\doibase
  https://doi.org/10.1016/j.aop.2015.10.010} {\bibfield  {journal} {\bibinfo
  {journal} {Annals of Physics}\ }\textbf {\bibinfo {volume} {363}},\ \bibinfo
  {pages} {476 } (\bibinfo {year} {2015})}\BibitemShut {NoStop}%
\bibitem [{\citenamefont {da~Silva}\ \emph {et~al.}(2009)\citenamefont
  {da~Silva}, \citenamefont {Gleria}, \citenamefont {Lyra},\ and\ \citenamefont
  {Sombra}}]{daSilva:09}%
  \BibitemOpen
  \bibfield  {author} {\bibinfo {author} {\bibnamefont {da~Silva},
  \bibfnamefont {G.~L.}}, \bibinfo {author} {\bibnamefont {Gleria},
  \bibfnamefont {I.}}, \bibinfo {author} {\bibnamefont {Lyra}, \bibfnamefont
  {M.~L.}}, \ and\ \bibinfo {author} {\bibnamefont {Sombra}, \bibfnamefont
  {A.~S.~B.}},\ }\href {\doibase 10.1364/JOSAB.26.000183} {\bibfield  {journal}
  {\bibinfo  {journal} {J. Opt. Soc. Am. B}\ }\textbf {\bibinfo {volume}
  {26}},\ \bibinfo {pages} {183} (\bibinfo {year} {2009})}\BibitemShut
  {NoStop}%
\bibitem [{\citenamefont {da~Silva}, \citenamefont {Lobo},\ and\ \citenamefont
  {Canabarro}(2014)}]{canabarro3}%
  \BibitemOpen
  \bibfield  {author} {\bibinfo {author} {\bibnamefont {da~Silva},
  \bibfnamefont {G.~L.}}, \bibinfo {author} {\bibnamefont {Lobo}, \bibfnamefont
  {T.~P.}}, \ and\ \bibinfo {author} {\bibnamefont {Canabarro}, \bibfnamefont
  {A.~A.}},\ }\href {\doibase 10.1364/JOSAB.31.002012} {\bibfield  {journal}
  {\bibinfo  {journal} {J. Opt. Soc. Am. B}\ }\textbf {\bibinfo {volume}
  {31}},\ \bibinfo {pages} {2012} (\bibinfo {year} {2014})}\BibitemShut
  {NoStop}%
\bibitem [{\citenamefont {Solli}\ \emph {et~al.}(2007)\citenamefont {Solli},
  \citenamefont {Ropers}, \citenamefont {Koonath},\ and\ \citenamefont
  {Jalali}}]{Solli2007}%
  \BibitemOpen
  \bibfield  {author} {\bibinfo {author} {\bibnamefont {Solli}, \bibfnamefont
  {D.~R.}}, \bibinfo {author} {\bibnamefont {Ropers}, \bibfnamefont {C.}},
  \bibinfo {author} {\bibnamefont {Koonath}, \bibfnamefont {P.}}, \ and\
  \bibinfo {author} {\bibnamefont {Jalali}, \bibfnamefont {B.}},\ }\href
  {\doibase 10.1038/nature06402} {\bibfield  {journal} {\bibinfo  {journal}
  {Nature}\ }\textbf {\bibinfo {volume} {450}},\ \bibinfo {pages} {1054}
  (\bibinfo {year} {2007})}\BibitemShut {NoStop}%
\bibitem [{\citenamefont {Tanemura}\ and\ \citenamefont
  {Kikuchi}(2003)}]{Tanemura03}%
  \BibitemOpen
  \bibfield  {author} {\bibinfo {author} {\bibnamefont {Tanemura},
  \bibfnamefont {T.}}\ and\ \bibinfo {author} {\bibnamefont {Kikuchi},
  \bibfnamefont {K.}},\ }\href {\doibase 10.1364/JOSAB.20.002502} {\bibfield
  {journal} {\bibinfo  {journal} {J. Opt. Soc. Am. B}\ }\textbf {\bibinfo
  {volume} {20}},\ \bibinfo {pages} {2502} (\bibinfo {year}
  {2003})}\BibitemShut {NoStop}%
\bibitem [{\citenamefont {Trillo}\ \emph {et~al.}(1989)\citenamefont {Trillo},
  \citenamefont {Wabnitz}, \citenamefont {Stegeman},\ and\ \citenamefont
  {Wright}}]{Trillo89}%
  \BibitemOpen
  \bibfield  {author} {\bibinfo {author} {\bibnamefont {Trillo}, \bibfnamefont
  {S.}}, \bibinfo {author} {\bibnamefont {Wabnitz}, \bibfnamefont {S.}},
  \bibinfo {author} {\bibnamefont {Stegeman}, \bibfnamefont {G.~I.}}, \ and\
  \bibinfo {author} {\bibnamefont {Wright}, \bibfnamefont {E.~M.}},\ }\href
  {\doibase 10.1364/JOSAB.6.000889} {\bibfield  {journal} {\bibinfo  {journal}
  {J. Opt. Soc. Am. B}\ }\textbf {\bibinfo {volume} {6}},\ \bibinfo {pages}
  {889} (\bibinfo {year} {1989})}\BibitemShut {NoStop}%
\bibitem [{\citenamefont {Wu}\ and\ \citenamefont {Kalinikos}(2008)}]{wuPRL}%
  \BibitemOpen
  \bibfield  {author} {\bibinfo {author} {\bibnamefont {Wu}, \bibfnamefont
  {M.}}\ and\ \bibinfo {author} {\bibnamefont {Kalinikos}, \bibfnamefont
  {B.~A.}},\ }\href {\doibase 10.1103/PhysRevLett.101.027206} {\bibfield
  {journal} {\bibinfo  {journal} {Phys. Rev. Lett.}\ }\textbf {\bibinfo
  {volume} {101}},\ \bibinfo {pages} {027206} (\bibinfo {year}
  {2008})}\BibitemShut {NoStop}%
\bibitem [{\citenamefont {Xiang}\ \emph {et~al.}(2011)\citenamefont {Xiang},
  \citenamefont {Dai}, \citenamefont {Wen},\ and\ \citenamefont
  {Fan}}]{Xiang11}%
  \BibitemOpen
  \bibfield  {author} {\bibinfo {author} {\bibnamefont {Xiang}, \bibfnamefont
  {Y.}}, \bibinfo {author} {\bibnamefont {Dai}, \bibfnamefont {X.}}, \bibinfo
  {author} {\bibnamefont {Wen}, \bibfnamefont {S.}}, \ and\ \bibinfo {author}
  {\bibnamefont {Fan}, \bibfnamefont {D.}},\ }\href {\doibase
  10.1364/JOSAB.28.000908} {\bibfield  {journal} {\bibinfo  {journal} {J. Opt.
  Soc. Am. B}\ }\textbf {\bibinfo {volume} {28}},\ \bibinfo {pages} {908}
  (\bibinfo {year} {2011})}\BibitemShut {NoStop}%
\bibitem [{\citenamefont {Zakharov}\ and\ \citenamefont
  {Ostrovsky}(2009)}]{ZAKHAROV2009540}%
  \BibitemOpen
  \bibfield  {author} {\bibinfo {author} {\bibnamefont {Zakharov},
  \bibfnamefont {V.}}\ and\ \bibinfo {author} {\bibnamefont {Ostrovsky},
  \bibfnamefont {L.}},\ }\href {\doibase
  https://doi.org/10.1016/j.physd.2008.12.002} {\bibfield  {journal} {\bibinfo
  {journal} {Physica D: Nonlinear Phenomena}\ }\textbf {\bibinfo {volume}
  {238}},\ \bibinfo {pages} {540 } (\bibinfo {year} {2009})}\BibitemShut
  {NoStop}%
\bibitem [{\citenamefont {Zhang}\ \emph {et~al.}(2014)\citenamefont {Zhang},
  \citenamefont {Xiang}, \citenamefont {Dai},\ and\ \citenamefont
  {Wen}}]{Zhang14}%
  \BibitemOpen
  \bibfield  {author} {\bibinfo {author} {\bibnamefont {Zhang}, \bibfnamefont
  {L.}}, \bibinfo {author} {\bibnamefont {Xiang}, \bibfnamefont {Y.}}, \bibinfo
  {author} {\bibnamefont {Dai}, \bibfnamefont {X.}}, \ and\ \bibinfo {author}
  {\bibnamefont {Wen}, \bibfnamefont {S.}},\ }\href {\doibase
  10.1364/JOSAB.31.003029} {\bibfield  {journal} {\bibinfo  {journal} {J. Opt.
  Soc. Am. B}\ }\textbf {\bibinfo {volume} {31}},\ \bibinfo {pages} {3029}
  (\bibinfo {year} {2014})}\BibitemShut {NoStop}%
\bibitem [{\citenamefont {Zhong}, \citenamefont {Cheng},\ and\ \citenamefont
  {Chiang}(2014)}]{Zhong14}%
  \BibitemOpen
  \bibfield  {author} {\bibinfo {author} {\bibnamefont {Zhong}, \bibfnamefont
  {X.}}, \bibinfo {author} {\bibnamefont {Cheng}, \bibfnamefont {K.}}, \ and\
  \bibinfo {author} {\bibnamefont {Chiang}, \bibfnamefont {K.~S.}},\ }\href
  {\doibase 10.1364/JOSAB.31.001484} {\bibfield  {journal} {\bibinfo  {journal}
  {J. Opt. Soc. Am. B}\ }\textbf {\bibinfo {volume} {31}},\ \bibinfo {pages}
  {1484} (\bibinfo {year} {2014})}\BibitemShut {NoStop}%
\bibitem [{\citenamefont {Zhong}\ and\ \citenamefont
  {Xiang}(2007)}]{ZHONG2007271}%
  \BibitemOpen
  \bibfield  {author} {\bibinfo {author} {\bibnamefont {Zhong}, \bibfnamefont
  {X.}}\ and\ \bibinfo {author} {\bibnamefont {Xiang}, \bibfnamefont {A.}},\
  }\href {\doibase https://doi.org/10.1016/j.yofte.2007.04.001} {\bibfield
  {journal} {\bibinfo  {journal} {Optical Fiber Technology}\ }\textbf {\bibinfo
  {volume} {13}},\ \bibinfo {pages} {271 } (\bibinfo {year}
  {2007})}\BibitemShut {NoStop}%
\bibitem [{\citenamefont {Zhou}\ \emph {et~al.}(2009)\citenamefont {Zhou},
  \citenamefont {Su}, \citenamefont {Cheng}, \citenamefont {Xiang},
  \citenamefont {Dai},\ and\ \citenamefont {Wen}}]{ZHOU20091440}%
  \BibitemOpen
  \bibfield  {author} {\bibinfo {author} {\bibnamefont {Zhou}, \bibfnamefont
  {W.}}, \bibinfo {author} {\bibnamefont {Su}, \bibfnamefont {W.}}, \bibinfo
  {author} {\bibnamefont {Cheng}, \bibfnamefont {X.}}, \bibinfo {author}
  {\bibnamefont {Xiang}, \bibfnamefont {Y.}}, \bibinfo {author} {\bibnamefont
  {Dai}, \bibfnamefont {X.}}, \ and\ \bibinfo {author} {\bibnamefont {Wen},
  \bibfnamefont {S.}},\ }\href {\doibase
  https://doi.org/10.1016/j.optcom.2008.12.037} {\bibfield  {journal} {\bibinfo
   {journal} {Optics Communications}\ }\textbf {\bibinfo {volume} {282}},\
  \bibinfo {pages} {1440 } (\bibinfo {year} {2009})}\BibitemShut {NoStop}%
\bibitem [{\citenamefont {Ziolkowski}(1997)}]{debye2}%
  \BibitemOpen
  \bibfield  {author} {\bibinfo {author} {\bibnamefont {Ziolkowski},
  \bibfnamefont {R.~W.}},\ }\href {\doibase 10.1109/8.558653} {\bibfield
  {journal} {\bibinfo  {journal} {IEEE Transactions on Antennas and
  Propagation}\ }\textbf {\bibinfo {volume} {45}},\ \bibinfo {pages} {375}
  (\bibinfo {year} {1997})}\BibitemShut {NoStop}%
\bibitem [{\citenamefont {Ziolkowski}\ and\ \citenamefont
  {Judkins}(1993)}]{Ziolkowski:93}%
  \BibitemOpen
  \bibfield  {author} {\bibinfo {author} {\bibnamefont {Ziolkowski},
  \bibfnamefont {R.~W.}}\ and\ \bibinfo {author} {\bibnamefont {Judkins},
  \bibfnamefont {J.~B.}},\ }\href {\doibase 10.1364/JOSAB.10.000186} {\bibfield
   {journal} {\bibinfo  {journal} {J. Opt. Soc. Am. B}\ }\textbf {\bibinfo
  {volume} {10}},\ \bibinfo {pages} {186} (\bibinfo {year} {1993})}\BibitemShut
  {NoStop}%
\end{thebibliography}%

\end{document}